\documentclass{aastex63}
\usepackage{amsmath}
\usepackage{amsfonts}
\usepackage{amssymb}
\usepackage{multirow}
\usepackage{graphicx}

\def \apjs {{\rm {ApJS}}}

\def \aap {{\rm {A\&A}}}

\def \apjl{\rm {ApJL}}

\newcommand{\ms}{\mbox{\,m s$^{-1}$}}
\newcommand{\msyr}{\mbox{\,m s$^{-1}$yr$^{-1}$}}
\newcommand{\cms}{\mbox{\,cm s$^{-1}~$}}
\date{\today}
\received{xxx}
\revised{xxx}
\accepted{xxx}
\submitjournal{ApJS}

\shorttitle{Search for nearby Earth analogs}
\shortauthors{Feng et al.}

\begin{document}
\title{Search for Nearby Earth Analogs.\\III. Detection of ten new planets, three planet candidates, and confirmation of three planets around eleven nearby M dwarfs}
\author[0000-0001-6039-0555]{Fabo Feng}
\affiliation{Earth and Planets Laboratory, Carnegie Institution for Science, Washington, DC 20015, USA}
\author{Stephen A. Shectman}
\affiliation{Observatories of the Carnegie Institution for Science, 813 Santa Barbara St., Pasadena, CA 91101}
\author{Matthew S. Clement}
\affiliation{Earth and Planets Laboratory, Carnegie Institution for Science, Washington, DC 20015, USA}
\author{Steven S. Vogt}
\affiliation{UCO/Lick Observatory, University of California, Santa
  Cruz, CA 95064,USA}
\author{Mikko Tuomi}
\affiliation{ Centre for Astrophysics Research, University of Hertfordshire, Colllege Lane, AL10 9AB, Hatfield, UK}
\author{Johanna K. Teske}
\affiliation{Earth and Planets Laboratory, Carnegie Institution for Science, Washington, DC 20015, USA}
\affiliation{Observatories of the Carnegie Institution for Science, 813 Santa Barbara St., Pasadena, CA 91101}
\affiliation{Hubble Fellow}
\author{Jennifer Burt}
\affiliation{Jet Propulsion Laboratory, California Institute of Technology, 4800 Oak Grove drive, Pasadena CA 91109}
\author{Jeffrey D. Crane}
\affiliation{Observatories of the Carnegie Institution for Science, 813 Santa Barbara St., Pasadena, CA 91101}
\author{Bradford Holden}
\affiliation{UCO/Lick Observatory, University of California, Santa
  Cruz, CA 95064,USA}
\author{Sharon Xuesong Wang}
\affiliation{Observatories of the Carnegie Institution for Science, 813 Santa Barbara St., Pasadena, CA 91101}
\author{Ian B. Thompson}
\affiliation{Observatories of the Carnegie Institution for Science,
  813 Santa Barbara St., Pasadena, CA 91101}
\author{Mat\'ias R. D\'iaz}
\affiliation{Observatories of the Carnegie Institution for Science, 813 Santa Barbara St., Pasadena, CA 91101}
\affiliation{Departamento de Astronom\'ia, Universidad de Chile, Camino El Observatorio 1515, Las Condes, Santiago, Chile}
\author{R. Paul Butler}
\affiliation{Earth and Planets Laboratory, Carnegie Institution for Science, Washington, DC 20015, USA}

\correspondingauthor{Fabo Feng}
\email{fengfabo@gmail.com}

\begin{abstract}
Earth-sized planets in the habitable zones of M dwarfs are good
candidates for the study of habitability and detection of
biosignatures. To search for these planets, we analyze all available
radial velocity data and apply four signal detection criteria to
select the optimal candidates. We find ten strong candidates
satisfying these criteria and three weak candidates showing
inconsistency over time due to data samplings. We also confirm three previous planet candidates and
improve their orbital solutions through combined analyses of updated
data sets. Among the strong planet candidates, HIP 38594 b is a
temperate super-Earth with a mass of $8.2\pm 1.7$\,$M_\oplus$ and an orbital period of 60.7$\pm$0.1\,days, orbiting around an
early-type M dwarf. Early-type M dwarfs are less active and thus are
better hosts for habitable planets than mid-type and late-type M
dwarfs. Moreover, we report the detection of five two-planet systems,
including two systems made up of a warm or cold Neptune and a cold Jupiter, consistent with a
positive correlation between these two types of planets. We also
detect three temperate Neptunes, four cold Neptunes, and four cold
Jupiters, contributing to a rarely explored planet population. Due to their proximity to the
Sun, these planets on wide orbits are appropriate targets for direct
imaging by future facilities such as HabEx and ELT. 
\end{abstract}
\keywords{Exoplanet astronomy (486), Radial velocity (1332), Exoplanet
  detection methods (489), M dwarf stars (982), Astrostatistics
  (1882), High resolution spectroscopy (2096)}

\section{Introduction}\label{sec:intro}
One of the fundamental questions to humanity is whether there are other habitable worlds like the Earth.  Since Earth is the only planet known to host life, we imagine that the best candidates for habitable worlds are Earth-sized planets around Sun-like stars (though there is currently no data to confirm this bias). However, the Earth only induces 0.09\ms
radial velocity (RV) variation on the Sun and $\sim$84 parts per million
(ppm) transit depth. Signals with such a small transit and such as
long period are beyond the capabilities of any existing advanced instrument/telescope.
Modern facilities are sensitive to Earth-sized planets around low-mass stars (so-called Earth analogs)
such as M dwarfs. Although M dwarfs are more active than Sun-like
stars and the planets in their habitable zones (HZs;
\citealt{kopparapu14}) presumably evolved to possess tidally locked
synchronous orbits, there are plausible mechanisms to reduce the harm caused by stellar flaring
and tidal locking \citep{tarter07,shields16}. As $\sim$70\% of the
stars in our Galaxy are M dwarfs according to the RECON sample of
nearest stars\footnote{\url{http://www.recons.org/census.posted.htm}},
the Earth-sized planets around these low mass stars provide an
important sample for habitability studies and biosignature
searches.

On the other hand, because early-type M dwarfs are less active than mid-type and late-type M dwarfs
\citep{mohanty03,west15}, an HZ planet would require a weaker magnetic field to shield its planetary atmosphere from erosion by stellar activity
such as coronal mass ejections \citep{kay16}. Early-type M dwarfs are
also more abundant than Sun-like stars and have larger HZs and less
activity than other types of M dwarfs \citep{cuntz16,heller14}. Hence we could call early-type M
dwarfs ``Goldilocks M dwarfs'' for the search of habitable worlds. 

To date, the transit and RV methods have been used to discover about 20 Earth-sized HZ planets
around M dwarfs. Most of these temperate worlds are around
late-type M dwarfs, such as Proxima b, \citep{anglada16}, Teegarden's Star
b \citep{zechmeister19}, and the TRAPPIST-1 system
\citep{gillon17}. To increase the sample of Earth analogs, the RV
community has collected precision RV data for a few decades using
spectrometers such as the High Accuracy Radial velocity Planet
Searcher (HARPS; \citealt{pepe02}), the Planet Finder Spectrograph
\citep{crane06,crane08,crane10}, and the High Resolution Echelle Spectrometer mounted
on a KECK telescope (HIRES/KECK; \citealt{vogt94}). In
  particular, many infrared spectrographs including CARMENES
  \citep{quirrenbach09}, IRD \citep{tamura12}, HPF
  \citep{mahadevan12}, and SPIRou \citep{artigau14} are designed to be
  sensitive to Earth analogs around M dwarfs. The next generation high
  precision spectrographs such as ESPRESSO \citep{pepe10}, EXPRES
  \citep{jurgenson16}, and NEID \citep{schwab16} are able to improve
  the RV prevision to sub-m\,s$^{-1}$ level, marginally sensitive to Earth twins. While these instruments lay the foundation for extreme precision RV, multiple barriers must be overcomed to firmly detect signals caused by Earth analogs in noisy RV data.

Detection of Earth analogs is challenged by instrumental
instability (e.g., \citealt{halverson16} and \citealt{bechter18}),
stellar activity (e.g., \citealt{dumusque14} and \citealt{fischer16}),
and biased barycentric correction (e.g., \citealt{wright14} and
\citealt{feng19c}). In order to improve the efficiency and reliability
of the RV method, \cite{feng17b} developed the Agatha software suite to provide
comprehensive activity diagnoses. Moreover, to improve the barycentric
correction precision to 1\cms level, \cite{feng19c} created the PEXO
software to correctly model both the Earth’s barycentric motion and
the reflex motion of the target star by accounting for relativistic
effects. Recently our group developed an automated Agatha pipeline which has already been used to efficiently detect more than
20 planet candidates in \citealt{feng19a} (or paper I) and \citealt{feng20a} (or
paper II). In paper II, we reported two temperate super-Earths orbiting around
early-type M stars, indicating a large population of temperate worlds
embedded in the archived RV data. In this work, we continue to use our
automated pipeline to search for nearby Earth analogs around M dwarfs,
especially early-type ones. 

The paper is structured as follows. We introduce the RV data sets used
in this work in section \ref{sec:data} and briefly describe our
methodology in section \ref{sec:method}. Then we report the planet
candidates in section \ref{sec:planet} and study their dynamical
stability in section \ref{sec:stability}. Finally, we conclude in
section \ref{sec:conclusion}. 

\section{Data}\label{sec:data}
We select M dwarfs with RV data sets from the Automated Planet Finder (APF; \citealt{vogt14}), HARPS, HIRES/KECK, PFS, and SOPHIE \citep{perruchot08}. Based on
comprehensive analyses, we identify eleven stars that probably host
planets. The physical parameters and the number of RVs in each data set for each star are
shown in Table \ref{tab:par}. We use HARPSpre and HARPSpost to denote
the RV sets obtained before and after the fiber change for HARPS in
2015. Since we have not found discontinuity in RVs obtained before and
after upgrade of the PFS detector, we do not treat them independently
as we did in paper I and II. 
\begin{table}
\caption{Stellar parameters and information about RV data sets. The
  stellar type is given by the Simbad database \citep{wenger00}. The
  parallax is from {\it Gaia} DR2, the V magnitude is derived from the G
  magnitude from {\it Gaia} DR2 according to \cite{jordi10}, and the stellar mass is from TESS TIC
  input catalog \citep{stassun19}. To be simple, we use ``H1'' and ``H2'' respectively to denote
  HARPSpre and HARPSpost in this table. }
\label{tab:par}
\centering
\begin{tabular}{lllc cccc cccr}
  \hline\hline
Star & Other Name & Type & Stellar Mass & Parallax & V & APF & H1 & H2
  & KECK & PFS & SOPHIE\\
     &&&$M_\star$& mas & mag&&& &&&\\\hline
GJ 2056&HIP 34785&M0             &$0.62\pm0.08$&$35.13\pm0.03$&10.3&0&15&0&0&51&0 \\
GJ 317&LHS2037&M3.5V          &$0.42\pm0.02$&$65.77\pm0.06$&12&0&84&48&66&32&0 \\
GJ 480&HIP 61706&M3.5Ve         &$0.45\pm0.02$&$70.22\pm0.07$&11.5&0&37&0&21&0&0 \\
GJ 687&HIP 86162&M3.0V          &$0.40\pm0.02$&$219.78\pm0.03$&9.2&149&0&0&147&0&0 \\
GJ 9066&GJ 83.1&M4.5V          &$0.15\pm0.02$&$223.63\pm0.11$&12.5&0&25&0&54&0&0 \\
HIP 107772&TYC 7986-911-1&M0V            &$0.63\pm0.08$&$42.27\pm0.04$&10.5&0&22&0&0&49&0 \\
HIP 38594&TYC 6557-844-1&M0             &$0.61\pm0.02$&$56.19\pm0.03$&9.7&0&17&0&0&38&0 \\
HIP 4845&GJ 3072&M0V            &$0.62\pm0.04$&$47.37\pm0.04$&9.9&0&5&0&36&55&0 \\
HIP 48714&GJ 373&M0.5Ve         &$0.58\pm0.02$&$94.94\pm0.04$&8.9&119&0&0&22&0&12 \\
HIP 60559&Ross 695&M2             &$0.26\pm0.02$&$112.74\pm0.07$&11.3&0&24&0&17&0&0 \\
HIP 67164&GJ 3804&M3.5           &$0.34\pm0.02$&$89.23\pm0.08$&11.9&0&18&0&21&0&0 \\
\hline
\end{tabular}
\end{table}

As described in paper I and II, the KECK data was reduced and released by \cite{butler17}, and the PFS data are reduced using
the method developed by \cite{butler96} and \cite{butler06}. We use
the HARPS data reduced by \cite{trifonov20} using the SERVAL pipeline
\citep{zechmeister18}. The nightly RV zero points are subtracted from the
reduced data to reduce systematics. In addition to the data sets
used in paper II, we use the SOPHIE data released by \cite{soubiran18}
with correction of zero point drift \citep{courcol15}. Moreover, we use the RV data obtained by the Levy
spectrometer mounted on the 2.4\,m APF telescope. The APF data is reduced using the same pipeline as
used for PFS data reduction.

\begin{figure*}
  \centering
  \includegraphics[scale=0.6]{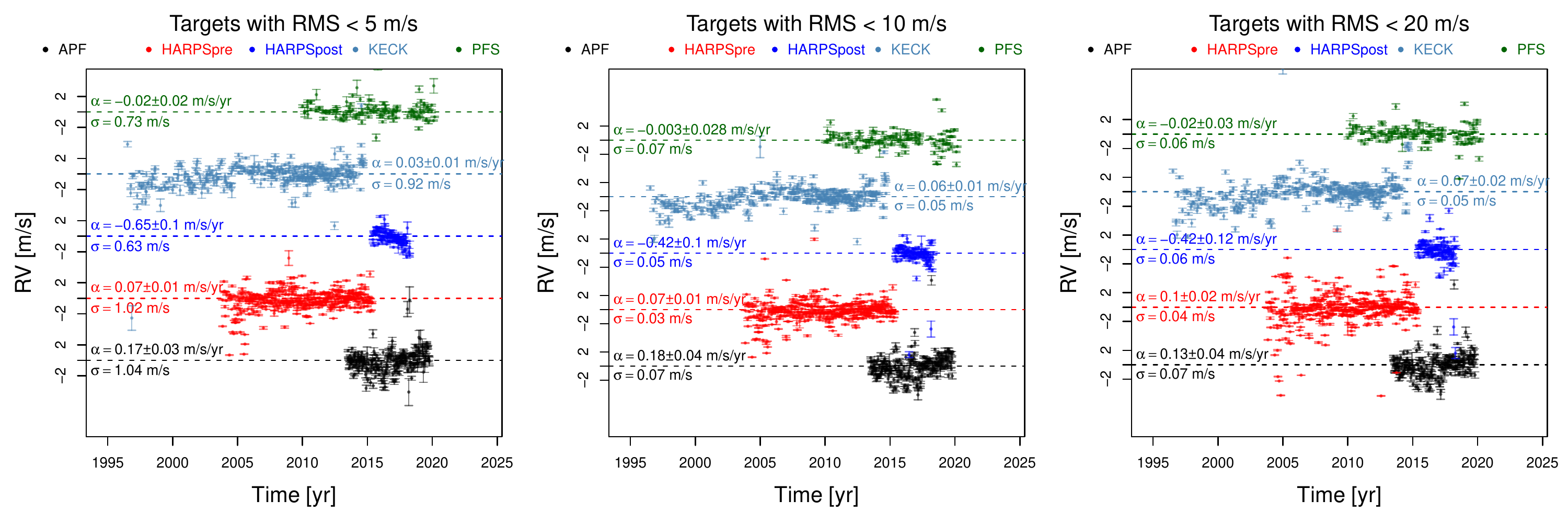}
  \caption{Comparison of the instrumental stability of APF,
    HARPSpre, HARPSpost, KECK, and PFS. The averaged RVs for different
    instruments are color-coded and shifted to optimize
    visualization. The slope ($\alpha$) and the standard deviation ($\sigma$) of residuals of the best-fit
    linear trend is shown for each instrument. }
  \label{fig:stability}
\end{figure*}

To compare the instrumental stabilities of APF (237 stars), PFS (573 stars), HARPSpre (2678 stars), HARPSpost
(917 stars), and KECK (1700 stars), we select the stars with more than
50 RVs and with standard deviations of less than 5, 10, and 20\ms  for each instrument. We bin the
data using a 10\,days time bin and calculate the weighted mean for
each instrument. We compare long term stability of these instruments through robust
linear regressions for the averaged data using the {\small R} package
{\small MASS} \citep{ripley13}. The weighted standard deviation ($\sigma$) of the residuals are
calculated using the {\small R} package {\small radiant.data} (\url{https://CRAN.R-project.org/package=radiant.data}). The
results are shown in Fig. \ref{fig:stability}. Since the residuals for
RMS$<$5\ms show much larger scattering than the ones selected by
higher RMS, we focus our investigation on the targets selected by RMS$<$10\ms, which is more conservative than RMS$<$20\ms.   

The slope for the best fit linear trend is 0.18$\pm$0.04\msyr for APF, 0.07$\pm$0.01\msyr for
HARPSpre, -0.42$\pm$0.10\msyr for HARPSpost, 0.06$\pm$0.01\msyr for
KECK, and 0.003$\pm$0.07\msyr for PFS. The HARPSpost and APF data show
the most significant linear trends, likely due to their shorter
observing baselines and smaller amount of RVs both of which make them
more sensitive to RV variations caused by planets and stellar
activity. On the other hand, the HARPSpre and KECK sets have longer
time spans and more RVs and thus are more suitable for stability
analysis. We find 0.06\msyr and 0.07\msyr accelerations for KECK and
HARPSpre at 6 and 7 sigma confidence levels, respectively. The
zero-point corrected KECK data \citep{talor19} also show a similar
linear trend with a slope of 0.05$\pm$0.01\msyr. Thus the
  zero-point correction is probably not able to remove long term
  bias in RV data. The linear trends are unlikely to be
  caused by outliers because they appear in the RV data selected by different critera. Moreover, the linear fits are weighted by measurement errors, reducing the influence of
  outliers which typically have large error bars. Hence the similar acceleration shared by KECK and HARPSpre suggests a common
astrophysical origin such as relativistic effects in the Solar System
as mentioned in \cite{feng19c}. There are also linear trends in APF
and HARPS data with a significance of about 4 sigma. Compared with the
other instruments, PFS is the most stable instrument over decade time
scales. The residuals for all instruments are less than 0.1\ms and
HARPSpre shows slightly better short term stability. The instability
bias in the data sets we use in this study is much less than 1\ms and
is thus negligible for the data analysis in this work. The SOPHIE data
has larger uncertainty even after drift correction and thus only plays a minor role in the constraints of orbits. Hence their instrumental instability is less important for this work. 

\section{Method}\label{sec:method}
\subsection{RV model}\label{sec:model}
Following paper I and II, we model the RV variation at epoch $t_j$ of data set $k$ induced by planets
using
\begin{equation}
  \hat{v}_{p,j}^k=\sum_{i=1}^{N_p}K_i\left[\sin{(\omega_i+\nu_i(t_j))}+e_i\cos{\omega_i}\right]+\gamma_k~,
\end{equation}
where $N_p$ is the number of planetary signals, $K_i$ is the
semi-amplitude of the RV variation induced by planet $i$, $\gamma_k$ is the
offset of RV set $k$, $\omega_i$ is the argument of periastron of
planet $i$, $e_i$ is eccentricity, $\nu_i(t_j)$ is the true anomaly of
planet $i$ at epoch $t_j$ and can be derived from the reference
mean anomaly $M_j(t_0)$ (or $M_{0,j}$), and period $P_j$ by solving
Kepler's equation. Here we do not use a linear trend to model
acceleration to avoid potential degeneracy between the linear trend and long period planet signals. 

We use the moving average (MA) model to account for time correlated
noise (or red noise) in RV data induced by stellar activity and
instrumental instability. Thus the full model for the RV at epoch
$t_j$ of set $k$ is
\begin{equation}
  \hat{v}_j^k=\hat{v}_{p,j}^k + \sum_{n=1}^{q}w_n^k\exp\left(-\frac{|t_j-t_{j-n}|}{\tau_k}\right)\left(v_{j-n}^k-\hat{v}_{p,j-n}^k\right)~,
\end{equation}
where $q$ is the order of MA model, $w_n^k$ is the amplitude of
MA component $n$ for set $k$, $\tau_k$ is the correlation time scale
for set $k$, $v_{j-n}^k$ is the observed RV at epoch $t_{j-n}$ of set $k$, and
$\hat{v}_{p,j-n}$ is the Keplerian RV at epoch $t_{j-n}$ of set
$k$. The MA model is found to be the so-called ``Goldilocks model'', which is able to
avoid false positives and false negatives according to the study of
synthetic and real RV sets \citep{feng16,ribas18} as well as the RV
fitting challenge \citep{dumusque16b}. Following \cite{feng17b}, we
compare different orders of MA models in the Bayesian
framework. Specifically, we select the highest order $q$, which passes the criterion that the
relative Bayesian Information Criterion
($\Delta$BIC, \citealt{spiegelhalter02}) of MA($q$) relative to MA($q-1$) is
larger than 10 \citep{kass95,feng16}. 

We model the excess noise in RV data using jitters in the logarithmic likelihood,
\begin{equation}
  {\rm ln}\mathcal{L}=-\frac{1}{2}\sum_{k=1}^{N_{\rm set}}\sum_{j=1}^{N_k}\left\{\ln\left[2\pi(\sigma_j^k+J_k^2)\right]+\frac{(v_j^k-\hat{v}_j^k)^2}{\sigma_j^k+J_k^2}\right\}~,
\end{equation}
where $J_k$ is the jitter for set $k$, $N_{\rm set}$ is the total
number of RV sets, $N_{k}$ is the number of epochs of set $k$. We
adopt logarithmic uniform priors for time scale parameters ($P$ and
$\tau$) and a semi-Gaussian prior, $P(e)=\mathcal{N}(\mu=0,\sigma=0.2)$, for eccentricity to capture the
broad feature of non-uniform eccentricity distributions found in
\cite{kane12} and \cite{VanEylen18}. The orbital solution is typically robust to
the change of $\sigma$ in the semi-Gaussian prior according to the
tests in paper II. We adopt uniform priors for other parameters. The boundary of these priors are
broad enough to allow all types of orbital solutions. 

To explore the posterior, we combine the adaptive Markov Chain Monte
Carlo (MCMC) developed by \cite{haario01} with a parallel scheme
developed in paper I. Specifically, we launch multiple tempered
(hot) chains (typically 16 chains) to explore the global posterior
maxima. Then untempered (cold) chains are launched to further sample
the global posterior maxima in order to find the optimal solution {\it a
  posteriori}. We start from 0-planet model and repeat these steps for
each additional planet until the decrease of BIC is less than
10. Considering that the posterior
distributions for multiple-planet systems are typically dominated by a
single Gaussian-like distribution, we follow \citet{kass95} and
\citet{feng16} to approximate the logarithmic Bayes factor (BF)
from $\Delta$BIC using
\begin{equation}
  {\rm ln}{\rm BF}\approx -\frac{1}{2}\Delta{\rm BIC}=\mathcal{L}_i^{\rm max}-\mathcal{L}_{i-1}^{\rm max}-\frac{1}{2}N_{\rm par}\ln{N_{\rm RV}}~,
\end{equation}
where $N_{\rm par}$ is the effective number of additional free parameters
by adding a Keplerian component onto $i-1$ Keplerian components, $N_{\rm RV}$ is the total number of
observed RVs. Thus the $\Delta{\rm BIC}>10$ criterion is equivalent to
$\ln{\rm BF}>5$. Since many planetary orbits are approximately circular,
eccentricity ($e$) and the argument of periastron ($\omega$) may not be
counted as effective as the other orbital parameters in terms of
improving the fitting. Thus we use lnBF$_3$ and lnBF$_5$ respectively to denote the lnBFs for a circular solution and an eccentric solution. We stop the MCMC samplers if lnBF$_3<5$.

\subsection{Signal diagnostics}\label{sec:diagnostics}
Following paper I and II, we diagnose whether an RV signal is related to stellar
activity or to planets using four criteria. First, a Keplerian signal should be
statistically significant. We regard signals passing ${\rm lnBF}_3>5$
as strong and ${\rm lnBF}_5>5$ as significant. In other words, the
former ones are strong candidates while the latter ones are weak candidates if they satisfy the other criteria as well.

Second, a Keplerian signal should be robust to the choice of noise
models. To implement this criterion, we calculate the Bayes factor
periodograms (BFPs, \citealt{feng17b}) for signals identified in a
combined RV set. We compare BFPs calculated using the MA(1) model (or
``MA'' for abbreviation), the first order autoregressive (AR(1) or ``AR'' for abbreviation; \citealt{tuomi13}) model and the
white noise (denoted by ``white'') model for each signal. The previous
signals are subtracted from the raw data for the calculation of BFPs for
subsequent signals. Unlike traditional periodograms, a BFP models excess white
noise using jitter and models correlated noise using red noise
models such as MA and AR. The default red noise model is MA, which is found to be appropriate for RV
data \citep{feng16}. For a given signal and a noise model, we
calculate the BF for each of a sample of periods by maximizing the likelihood using the
Marquardt-Levenberg algorithm \citep{levenberg44,marquardt63}. The
$\ln{\rm BF}>5$ criterion is then used as a threshold to assess the
statistical significance of a signal. However, the BFP is not suitable
for highly eccentric signals due to the assumption of circular
orbits. Hence we only use BFPs to test the sensitivity of signals to
noise models but rely on MCMCs to fully explore the posterior. 

Third, a Keplerian signal should not overlap with the periodic signals
found in activity indices or noise proxies. To search for activity
signals in noise proxies, we calculate BFPs for the sodium D lines,
H$\alpha$, the Ca II H and K lines and their corresponding S-index
measurements, along with the bisector and full width of half maximum
(FWHM) of spectral lines. We also count the window function as a noise proxy, which is used to exclude aliases and sampling
biases. If an RV signal is found to overlap with significant activity
signals, it is unlikely to be Keplerian. However, the chance of a
random Keplerian signal overlapping with signals in noise proxies is proportional to the number of proxies. In addition, long
period signals are more likely to overlap with activity signals because
their posterior are less constrained due to sparser sampling
of the orbit compared with short period signals. We typically adopt
$|P_{\rm rv}-P_{\rm act}|< 0.1P_{\rm rv}$ as a criterion to confirm significant overlap. 

Fourth, a Keplerian signal should be consistent over time. We calculate
the so-called moving periodogram (MP, \citealt{feng17b}) to implement
this criterion. To calculate MP, we define a time window and calculate
the BFP for the data covered by this time window. The MA(1) noise model is
used by default to account for red noise. We move the time window with a
certain time step and calculate the BFP for each step until the whole
time span is covered. The sequence of BFPs form a two dimensional
periodogram, called MP. If the signal is consistently significant over
time in MP, we regard it as time invariant and probably Keplerian. The time step is adjusted according
to the sampling and regularity of the combined data. For example, if
there is a great gap between two chunks of data, we may choose two time
windows and adjust time steps such that each window cover each chunk of
the data. An optimal time window should be several orbital periods of
a signal but also be small enough to select at most half of the
combined data. However, long period signals may not be appropriate for
MP because no time window is wide enough to cover one or two periods. A
rule of thumb is to calculate MP for signals with periods much shorter
than the data baseline ($T$), to set the window size to be $T/2$,
and to define a time step such that the time window covers the whole
baseline in 10 steps. To calculate the MPs consistently for different
data sets, we use a time window with a size of $T/2$ and a time step
of $T/20$ by default.

Finally, we combine the above four critera with eye inspection of the MP, the goodness of fit, and phase coverage of signals to diagnose and classify RV signals.

\section{Planet candidates}\label{sec:planet}

There is flexibility in the four criteria introduced in section
\ref{sec:diagnostics} for signal selection. For example, the criterion
of ln(BF)$>$5 depends on which number of efficient parameters one
chooses to calculate ln(BF). The MP criterion depends on the regularity of the data samplings and the
quality of the data. Thus we first select primordial signals which pass the
ln(BF$_3$)$>5$ criterion. We then investigate the origins of these signals by
checking the other criteria. In subsection \ref{sec:statistics}, we
classify the planet candidates into different categories according to
the four criteria and study the statistics of these new planet
candidates. In subsection \ref{sec:individual}, we discuss the results
for each target in detail. 

\subsection{Statistics of the new planet candidates}\label{sec:statistics}
The parameters of the planet candidates discovered in this work are
shown in Table \ref{tab:planet}. As in paper II, we classify the candidates into
different categories based on the detection criteria we have
introduced. There are ten strong candidates, three weak candidates, and
confirmation of three previous candidates. A strong candidate should
typically satisfy all of the four criteria. For candidates with
periods comparable with the data baseline, we don't apply the
time-consistency criterion because the MP is mainly designed to test
consistency of short period signals. A weak candidate does not satisfy some criteria due to legitimate reasons such as change in
significance caused by highly irregular RV sampling. We will discuss
individual cases in subsection \ref{sec:individual}. 

\begin{longrotatetable}
\hspace{-0.5in}
  \begin{deluxetable*}{llll llll lr}
\tablecaption{Parameters for the planet candidates. For each
  parameter, the value at the {\it maximum a posteriori} (MAP) and the uncertainty
  interval defined by the 1\% and 99\% quantiles of the posterior
  distribution are shown below the values of mean and standard
  deviation. $M_p\sin{I}$ is the minimum planet mass and $a$ is the semi-major axis. These two
  quantities are derived from the orbital period $P$, semi-amplitude
  $K$, and eccentricity $e$ by using the stellar masses given in Table
  \ref{tab:par}. $\omega$ is the argument of periastron, and $T_p$ is
  the epoch at the periastron. As introduced in section
  \ref{sec:method}, lnBF$_3$ is the logarithmic Bayes
  factor of the model including the candidate relative to the
  model excluding the candidate. Following paper II, we use the bold font to denote strong candidates, use the italic font to denote weak
  candidates, and use the normal font to denote confirmation of previous
  candidates. In the column of ``Note'', we fill ``HZ'' for candidates
  in the HZ, ``PHZ'' for candidates with orbits partially in the HZ,
  ``A12'' for candidates detected by \cite{anglada12b}, ``B14'' for candidates detected by \cite{burt14}, ``NC'' for candidates with inconsistent significance over time due to poor sampling, and ``PC'' for candidates with incomplete phase coverage.
   \label{tab:planet}}
  \tablehead{
\colhead{Planet}&\dcolhead{M_p\sin{I}}& \dcolhead{a} & \dcolhead{P}
&\dcolhead{K}&\dcolhead{e}&\dcolhead{\omega}&\dcolhead{T_p}&\colhead{$\ln{\rm BF}_3$}&\colhead{Note}\\
\colhead{} & \dcolhead{(M_\oplus)} & \colhead{(au)} & \colhead{(day)} & 
\colhead{(\ms)} & \colhead{} & \colhead{(deg)} &\colhead{(JD-2400000)} & \colhead{} & \colhead{}
}
\startdata
{\bf GJ 2056 b}&$16.2\pm3.6$&$0.283\pm0.013$&$69.971\pm0.061$&$5.23\pm1.51$&$0.72\pm0.10$&$58\pm116$&$52965.6\pm3.5$&10.4&HZ\\
&$17.6_{-8.6}^{+7.8}$&$0.283_{-0.033}^{+0.027}$&$69.937_{-0.164}^{+0.199}$&$4.89_{-2.29}^{+3.97}$&$0.64_{-0.17}^{+0.26}$&$13_{-12}^{+347}$&$52967.9_{-11.5}^{+9.1}$&&\\
{\it GJ 2056 c}&$141.2\pm17.0$&$3.453\pm0.164$&$2982.394\pm75.913$&$14.59\pm1.10$&$0.81\pm0.02$&$338\pm69$&$50505.3\pm156.1$&15.0&PC\\
&$141.2_{-39.0}^{+39.9}$&$3.469_{-0.436}^{+0.337}$&$2996.577_{-170.268}^{+178.294}$&$13.92_{-1.97}^{+2.97}$&$0.79_{-0.04}^{+0.06}$&$349_{-349}^{+11}$&$50472.0_{-378.0}^{+339.4}$&&\\
GJ 317 b&$557.1\pm18.3$&$1.151\pm0.018$&$695.660\pm0.355$&$71.81\pm0.58$&$0.07\pm0.01$&$72\pm133$&$50988.9\pm14.5$&256.2&A12\\
&$556.1_{-42.3}^{+43.3}$&$1.152_{-0.045}^{+0.040}$&$695.890_{-1.100}^{+0.619}$&$71.59_{-1.11}^{+1.81}$&$0.07_{-0.02}^{+0.02}$&$6_{-6}^{+353}$&$50986.2_{-29.2}^{+33.6}$&&\\
GJ 317 c&$522.5\pm19.1$&$5.230\pm0.111$&$6739.323\pm143.106$&$31.97\pm0.56$&$0.17\pm0.02$&$108\pm9$&$49804.6\pm221.5$&58.7&A12\\
&$529.0_{-52.9}^{+37.6}$&$5.223_{-0.251}^{+0.268}$&$6718.777_{-282.469}^{+375.180}$&$32.36_{-1.77}^{+0.84}$&$0.17_{-0.05}^{+0.05}$&$114_{-26}^{+18}$&$49916.0_{-611.2}^{+436.8}$&&\\
{\bf GJ 480 b}&$13.2\pm1.7$&$0.068\pm0.001$&$9.567\pm0.005$&$6.80\pm0.87$&$0.10\pm0.07$&$151\pm92$&$54562.1\pm2.3$&19.7&\\
&$13.5_{-4.4}^{+3.8}$&$0.068_{-0.003}^{+0.002}$&$9.565_{-0.009}^{+0.015}$&$6.92_{-2.20}^{+1.96}$&$0.04_{-0.03}^{+0.25}$&$118_{-114}^{+235}$&$54562.5_{-5.2}^{+4.0}$&&\\
GJ 687 b&$17.2\pm1.0$&$0.163\pm0.003$&$38.142\pm0.007$&$6.14\pm0.32$&$0.17\pm0.05$&$117\pm19$&$50592.7\pm2.0$&95.4&HZ,
B14\\
&$17.2_{-2.5}^{+2.4}$&$0.163_{-0.007}^{+0.006}$&$38.145_{-0.020}^{+0.015}$&$6.15_{-0.79}^{+0.67}$&$0.19_{-0.14}^{+0.10}$&$127_{-62}^{+34}$&$50593.6_{-6.6}^{+3.3}$&&\\
{\bf GJ 687 c}&$16.0\pm4.1$&$1.165\pm0.023$&$727.562\pm12.198$&$2.44\pm0.80$&$0.40\pm0.22$&$176\pm63$&$50304.1\pm292.8$&10.7&\\
&$19.4_{-12.4}^{+5.2}$&$1.164_{-0.054}^{+0.056}$&$726.403_{-29.204}^{+39.189}$&$3.69_{-2.74}^{+0.56}$&$0.72_{-0.71}^{+0.03}$&$189_{-178}^{+160}$&$49891.9_{-16.1}^{+710.3}$&&\\
{\bf GJ 9066 b}&$30.9\pm6.4$&$0.403\pm0.018$&$241.883\pm1.808$&$11.53\pm2.28$&$0.18\pm0.12$&$204\pm129$&$51273.4\pm50.7$&12.0&\\
&$27.9_{-11.0}^{+19.0}$&$0.403_{-0.047}^{+0.039}$&$241.590_{-3.996}^{+4.635}$&$10.29_{-3.73}^{+6.53}$&$0.16_{-0.16}^{+0.32}$&$5_{-3}^{+353}$&$51287.1_{-111.3}^{+119.4}$&&\\
{\bf GJ 9066 c}&$71.6\pm10.3$&$0.870\pm0.040$&$767.887\pm7.500$&$18.97\pm2.13$&$0.33\pm0.10$&$278\pm119$&$50884.0\pm53.3$&20.4&\\
&$70.7_{-21.5}^{+26.3}$&$0.871_{-0.100}^{+0.086}$&$766.954_{-19.237}^{+16.710}$&$18.98_{-4.81}^{+5.69}$&$0.39_{-0.32}^{+0.18}$&$331_{-330}^{+29}$&$50870.5_{-121.9}^{+148.8}$&&\\
{\bf HIP 107772 b}&$12.9\pm3.0$&$0.243\pm0.011$&$55.199\pm0.083$&$3.01\pm0.65$&$0.18\pm0.11$&$151\pm72$&$53199.6\pm11.2$&8.7&HZ\\
&$15.6_{-9.4}^{+4.6}$&$0.243_{-0.027}^{+0.022}$&$55.259_{-0.311}^{+0.111}$&$3.63_{-2.13}^{+0.96}$&$0.21_{-0.20}^{+0.25}$&$163_{-152}^{+184}$&$53199.0_{-22.4}^{+28.9}$&&\\
{\bf HIP 38594 b}&$8.1\pm1.7$&$0.256\pm0.003$&$60.722\pm0.122$&$1.89\pm0.40$&$0.17\pm0.11$&$125\pm100$&$52956.2\pm13.9$&6.2&HZ\\
&$8.2_{-4.3}^{+3.8}$&$0.256_{-0.007}^{+0.006}$&$60.711_{-0.192}^{+0.426}$&$1.90_{-1.00}^{+0.91}$&$0.19_{-0.18}^{+0.28}$&$42_{-39}^{+315}$&$52948.1_{-17.7}^{+40.8}$&&\\
{\bf HIP 38594 c}&$48.4\pm7.4$&$3.805\pm0.172$&$3477.768\pm229.001$&$2.89\pm0.42$&$0.16\pm0.10$&$196\pm68$&$51246.7\pm1337.9$&11.2&\\
&$42.9_{-11.2}^{+24.5}$&$3.842_{-0.441}^{+0.399}$&$3524.860_{-572.002}^{+540.850}$&$2.60_{-0.66}^{+1.31}$&$0.26_{-0.25}^{+0.16}$&$231_{-207}^{+114}$&$49725.1_{-457.2}^{+3247.8}$&&\\
{\bf HIP 4845 b}&$14.4\pm3.0$&$0.176\pm0.007$&$34.150\pm0.046$&$4.05\pm0.86$&$0.25\pm0.14$&$108\pm72$&$54702.0\pm5.6$&13.0&\\
&$16.7_{-9.0}^{+5.2}$&$0.177_{-0.020}^{+0.016}$&$34.151_{-0.090}^{+0.170}$&$4.79_{-2.61}^{+1.40}$&$0.36_{-0.35}^{+0.21}$&$87_{-79}^{+262}$&$54701.3_{-12.4}^{+19.2}$&&\\
{\bf HIP 48714 b}&$22.9\pm2.8$&$0.112\pm0.001$&$17.818\pm0.002$&$9.35\pm1.29$&$0.50\pm0.08$&$202\pm14$&$51539.8\pm1.0$&23.4&\\
&$22.5_{-6.3}^{+7.2}$&$0.112_{-0.003}^{+0.003}$&$17.819_{-0.009}^{+0.004}$&$9.56_{-3.17}^{+2.99}$&$0.57_{-0.27}^{+0.11}$&$202_{-31}^{+37}$&$51539.6_{-1.9}^{+4.1}$&&\\
{\it HIP 60559 b}&$15.9\pm3.8$&$0.297\pm0.008$&$115.796\pm0.493$&$5.27\pm1.26$&$0.21\pm0.11$&$257\pm99$&$51474.3\pm32.8$&11.6&NC\\
&$17.1_{-8.9}^{+9.8}$&$0.298_{-0.019}^{+0.017}$&$115.873_{-1.585}^{+0.837}$&$5.70_{-3.05}^{+2.96}$&$0.28_{-0.27}^{+0.21}$&$299_{-297}^{+59}$&$51454.6_{-17.9}^{+96.8}$&&\\
{\it HIP 67164 b}&$9.7\pm2.3$&$0.067\pm0.001$&$10.942\pm0.004$&$6.06\pm1.62$&$0.21\pm0.16$&$175\pm76$&$51576.6\pm2.4$&12.3&NC\\
&$11.8_{-7.5}^{+3.6}$&$0.067_{-0.003}^{+0.003}$&$10.941_{-0.007}^{+0.008}$&$9.30_{-6.67}^{+1.35}$&$0.66_{-0.66}^{+0.02}$&$177_{-170}^{+176}$&$51576.8_{-5.5}^{+5.0}$&&\\
\enddata
  \end{deluxetable*}
\end{longrotatetable}

\begin{figure*}
  \centering
  \includegraphics[scale=0.4]{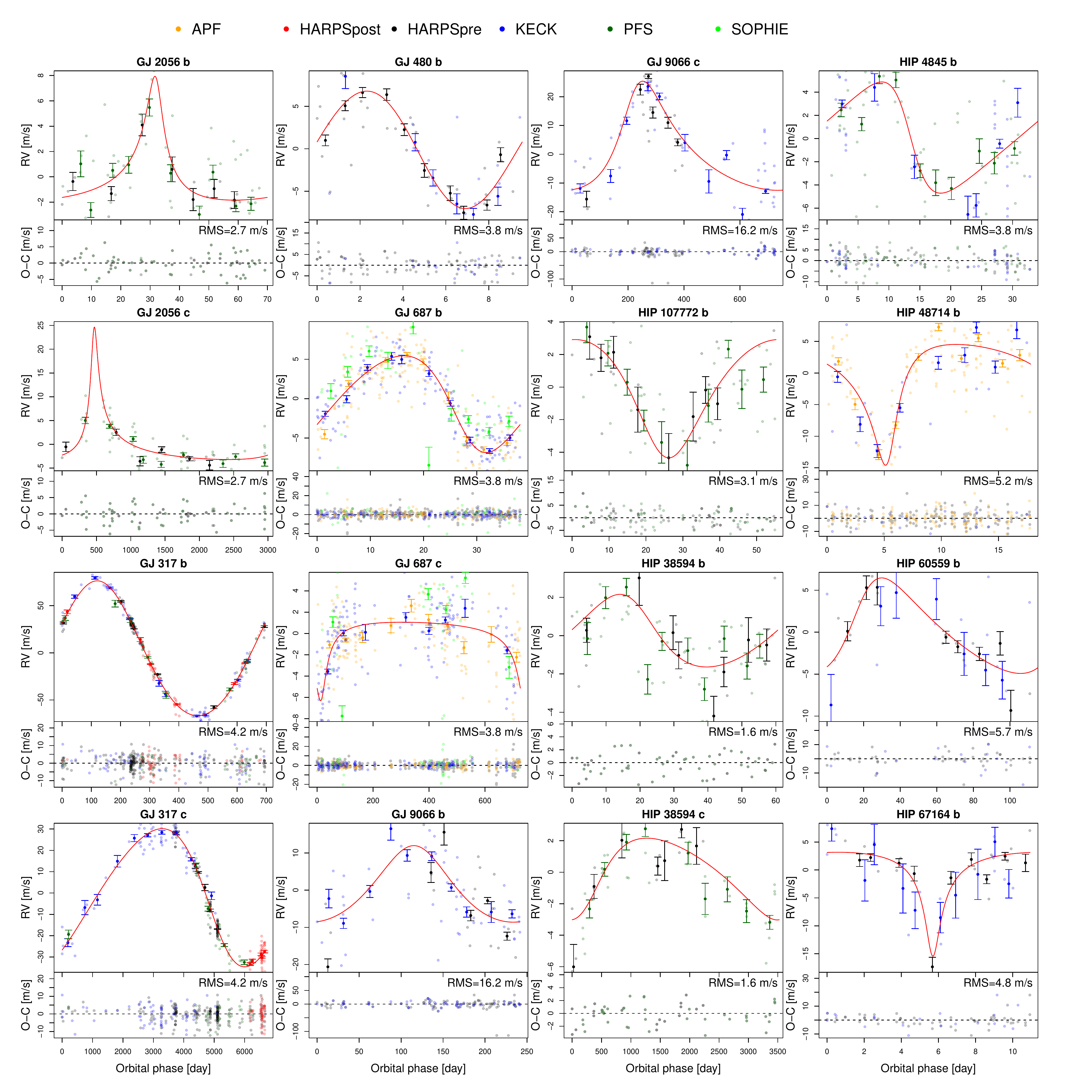}
  \caption{Phase curves and corresponding residuals for all planet candidates are shown. The instruments are encoded by different colors and
    are shown on top of all panels. The raw data measured by an instrument are binned using ten
    time bins that are regularly spaced over the orbital phase. The average and error of the RV for each bin are respectively calculated through weighted averages of the RVs and RV errors in each bin. The best orbital solution is determined by the MAP values of orbital parameters. The Root Mean Square (RMS) of the residual RVs
after subtracting all signals is shown in each panel.}
  \label{fig:phase}
\end{figure*}

\begin{figure*}
  \centering
  \includegraphics[scale=0.8]{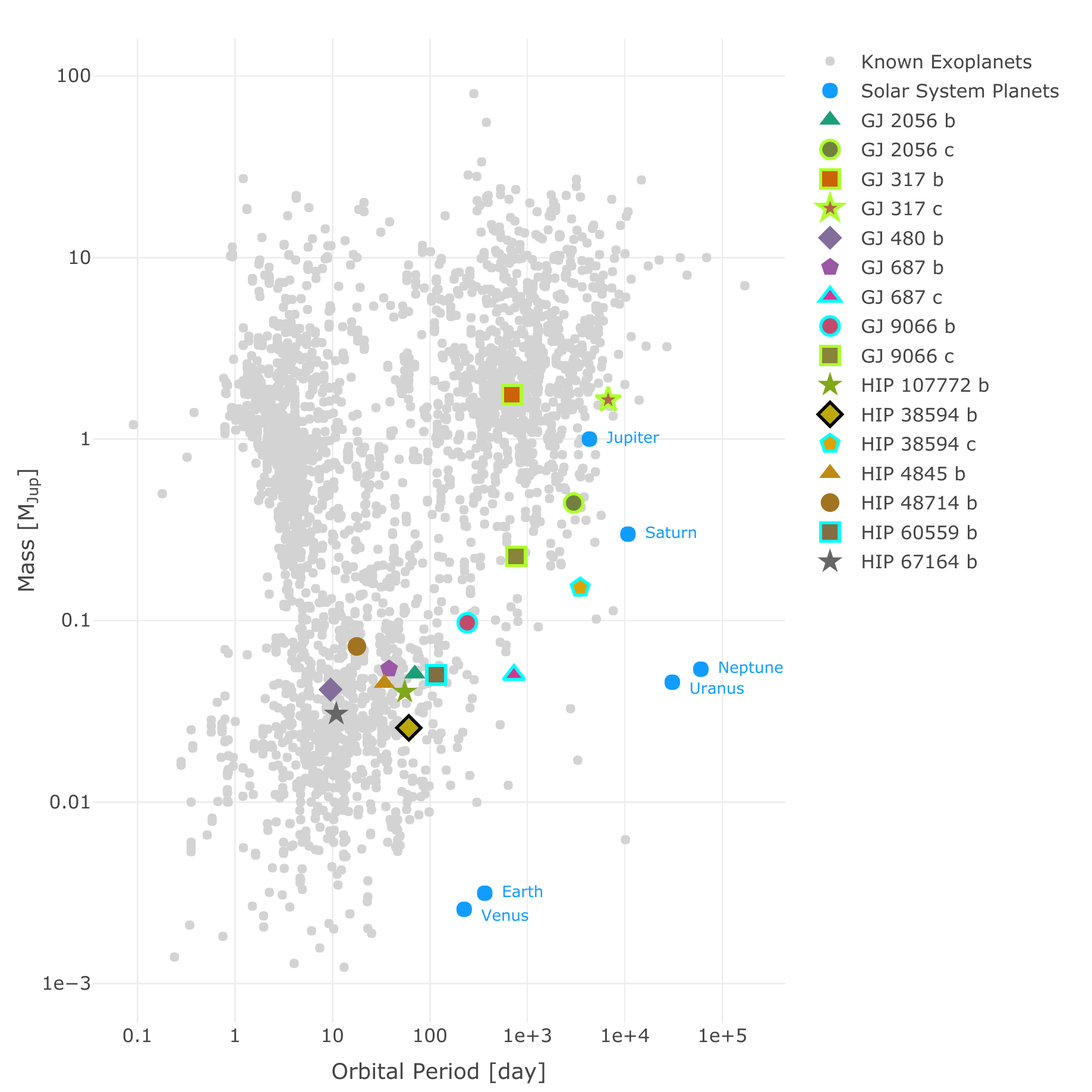}
  \caption{Mass and period distribution of known planets and the
    planet candidates found in this work. The candidates are
    represented by different shapes and colors of markers. HIP 38594 b
    is a temperate super-Earth, denoted by the black-edged diamond marker. The Jupiter analogs are denoted
    by the yellow-green-edged cross while the cyan-edged markers denote the
    warm and cold Neptunes detected in this work. The Solar System planets are
    represented by the blue dots. }
  \label{fig:dist}
\end{figure*}

We show the phase curves for these planet candidates in
Fig. \ref{fig:phase}. We also compare our candidates' mass and orbital
period distributions to those of other confirmed exoplanets in
Fig. \ref{fig:dist}. Thanks to the combined efforts of the RV community, we are able to detect cold
Neptunes with periods longer than 100\,days and minimum masses between
10 and 60\,$M_\oplus$, including GJ 687 c, GJ 9066 b, HIP
38594 c, and HIP 60559 b. Our detection of these cold Neptunes contributes
significantly to our understanding of this rarely explored
population in terms of increasing the sample size for studies of cold
Neptune formation. There are seven warm and cold Neptunes detected and two of
them also have cold Jupiter companions (GJ 2056 b and c, GJ 9066 b and
c). This is consistent with a positive correlation between super-Earths and
cold Jupiters as found by \cite{zhu18} and \cite{bryan19}, considering that
super-Earths and Neptunes are similar in size and mass. The
semi-amplitudes of these signals are as low as 1\ms, reaching the
limit of the current RV precision. Without further improvement of
instrument precision and stellar activity modeling, it seems to be
quite difficult to probe the $K<1$\ms regime although sub-m\,s$^{-1}$
signals have been reported for very stable and intensively observed
stars such as $\tau$ Ceti (e.g., \citealt{feng17c}). With longer observational baselines and larger compilations of high precision RV measurements, we believe that RV legacy data will play an important role to detect Jupiter analogs that
will be observed by CGI on WFIRST \citep{tang19}, MIRI on JWST \citep{danielski18}, the wide-field imager
MICADO mounted on the Extremely Large Telescope (ELT; \citealt{perrot18}) and HabEx \citep{gaudi20}. Such data will also be essential to confirm and
characterize the Jupiter analogs found by {\it Gaia} \citep{perryman14}. 

\begin{figure*}
  \centering
  \includegraphics[scale=1]{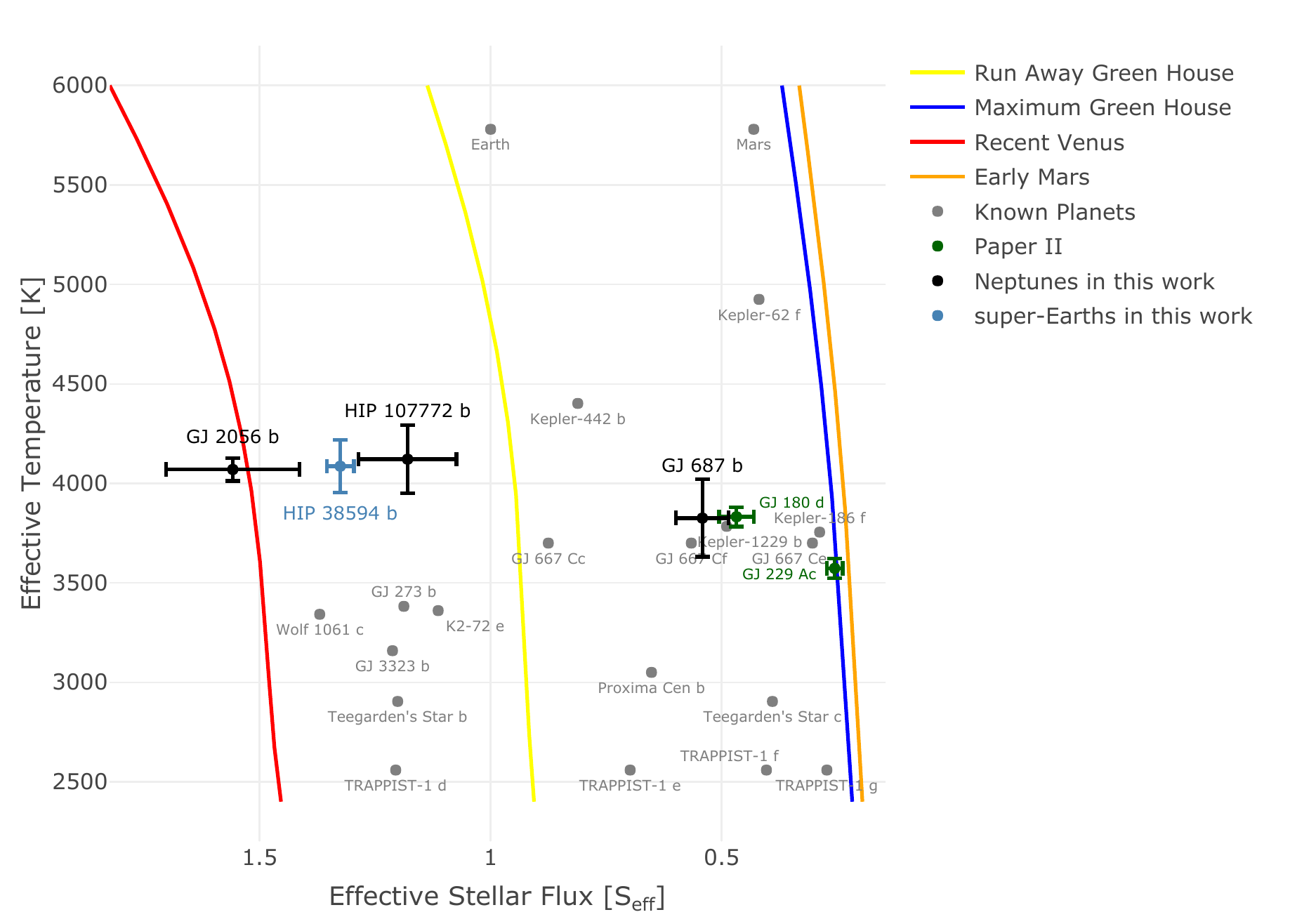}
  \caption{Known planets in the habitable zone. Based on Monte
    Carlo samplings of the effective stellar temperature and the
    stellar flux received by planets, we determine the error bars for
    the temperate planets detected in paper II (dark green), the HZ
    Neptunes (black), and the HZ super-Earth (light blue) found in this work. }
  \label{fig:hz}
\end{figure*}

Among the sample of candidates detected in this work, there is a
super-Earth candidate (HIP 38594 b) located in the optimistic HZ and
three temperate Neptunes (GJ 2056 b, HIP 107772 b, and GJ 687 b)
located in the HZ. These planets are shown in the context of the temperate super-Earths
reported in Paper II and other previously known HZ planets in
Fig. \ref{fig:hz}. Although the potential moons around these temperate
Neptunes might host liquid water on their surfaces, they are difficult
to detect given the current technology. Hence we focus our
investigation on HIP 38594 b, a temperate super-Earth. Compared with previous M dwarf hosts of temperate
planets, HIP 38594 is an early-type M dwarf and thus is less active in
terms of emitting energetic particles and UV light
\citep{mohanty03,west15}. Like K dwarfs, early-type M dwarfs are
Goldilocks stellar hosts because they are more abundant than Sun-like
stars and have a larger HZ than late-type M dwarfs
\citep{cuntz16,heller14}. HIP 38594 b is separated from HIP 38594 by
about 14\,mas and would thus make a promising target for direct imaging by ELT/MICADO \citep{perrot18}. 

\subsection{Individual planet candidates}\label{sec:individual}
\begin{figure*}
  \centering
  \includegraphics[scale=0.45]{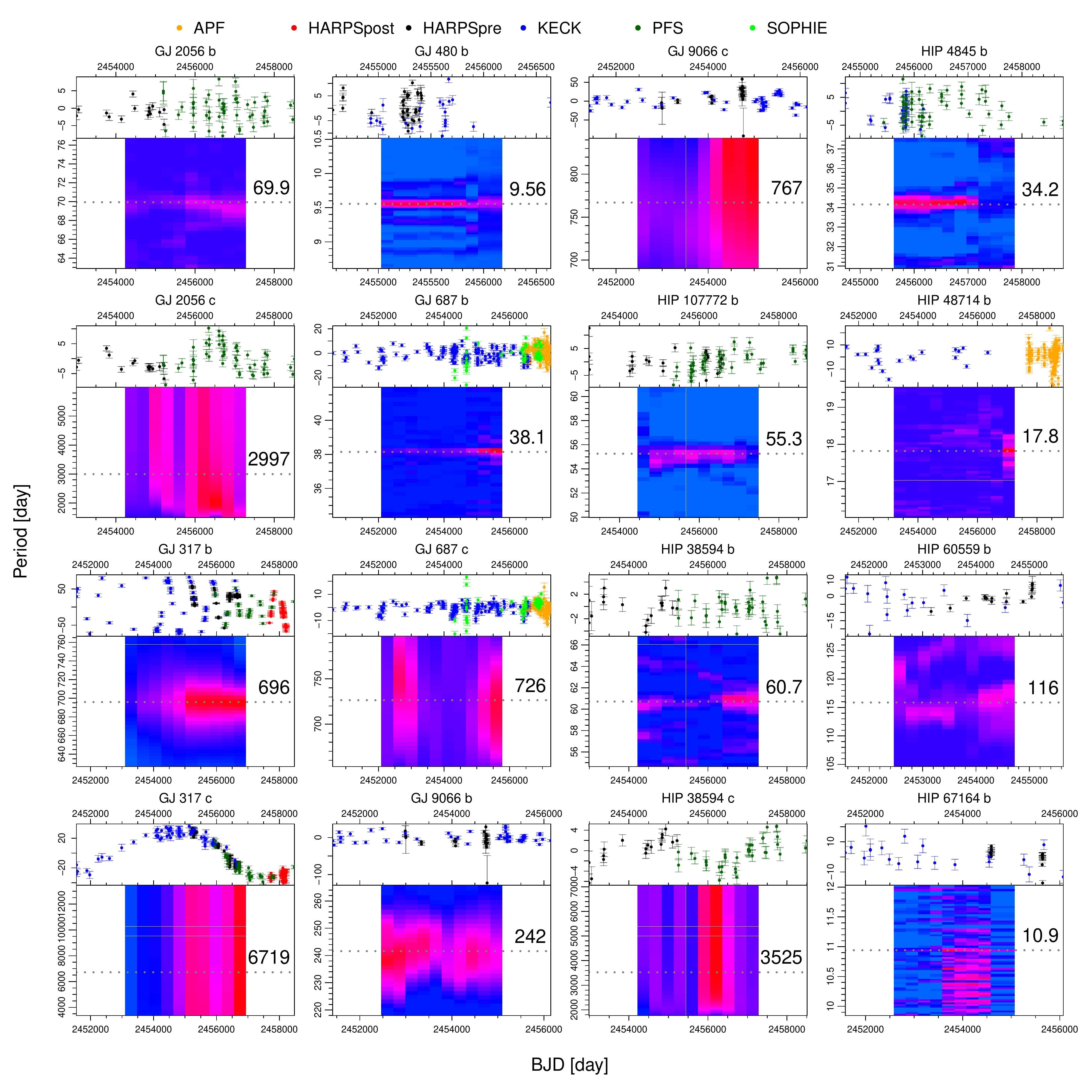}
  \caption{MPs for 16 signals. The width of the time window $\Delta T$
    is half of the time span ($T$) of the combined RV data set,
    i.e. $\delta T=T/2$. The time step is $\delta T=T/20$. It takes 20
    steps for the moving time window to cover the whole time span. For
  each plot, the upper panel shows the data minus the noise
  component and other RV signals. The optimal parameters of these model
  components are the MAP values. The lower panel is a zoom-in of the
  whole MP. The signal periods are denoted by horizontal dashed lines
and the MAP values in unit of days. }
  \label{fig:MP}
\end{figure*}

We discuss the results for individual targets by applying the
diagnostic criteria introduced in section \ref{sec:method}. The
MPs for all of the signals are shown in Fig. \ref{fig:MP}. We will use
these MPs as well as the BFPs for the RV data and the corresponding
activity indicators as clues for the investigation of the origin of
the identified signals. 

\begin{figure*}
  \centering
  \includegraphics[scale=0.5]{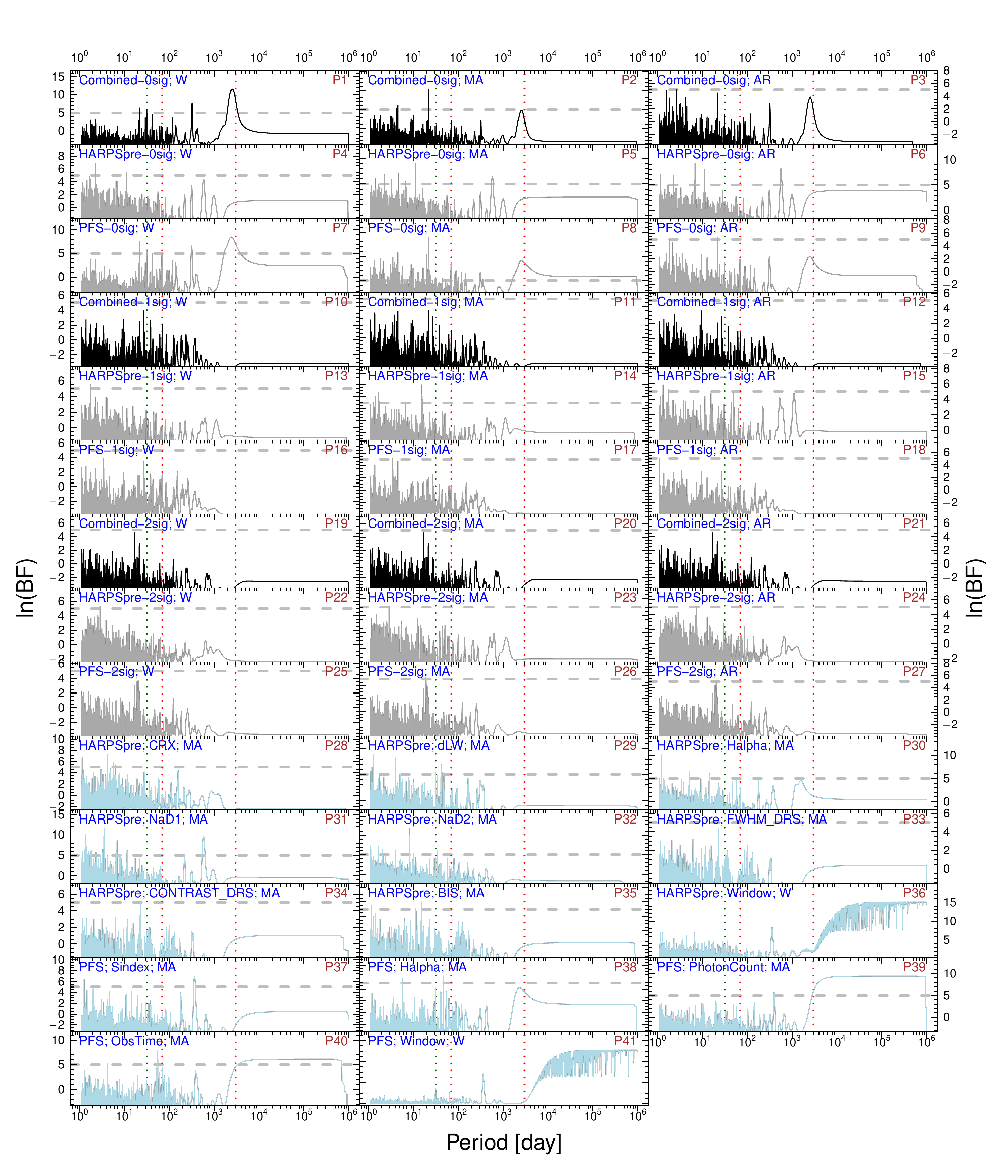}
  \caption{BFPs for the RVs and noise proxies for GJ 2056. The
    black BFPs are for the combined RV set with signals subtracted
    subsequently. The grey BFPs are for individual data sets, and the
    cyan BFPs are for noise proxies. The ln$(BF)=5$ threshold is denoted by
    the dashed lines. The window function is calculated using the
    Lomb-Scargle periodgoram \citep{lomb76,scargle82}. To be efficent,
    we use the generalized Lomb-Scargle periodogram with floating trend (GLST,
    \citealt{feng17b}) to show the signals in photometric data. The
    dark green dotted lines denote the rotation period of 32\,d from
    the literature. The red lines denote the two Keplerian signals at
    periods of 69.9 and 2997\,days. The data and noise model for a BFP is given in the top left corner. The white noise, MA, and AR models are dubbed by ``W'',
    ``MA'', and ``AR'' respectively. The panel number in each panel is
    shown for the reader to easily navigate the BFPs. The elements in
    subsequent BFP figures are defined in the same way.}
  \label{fig:GJ2056}
\end{figure*}

\begin{figure*}
  \centering
  \includegraphics[scale=0.45]{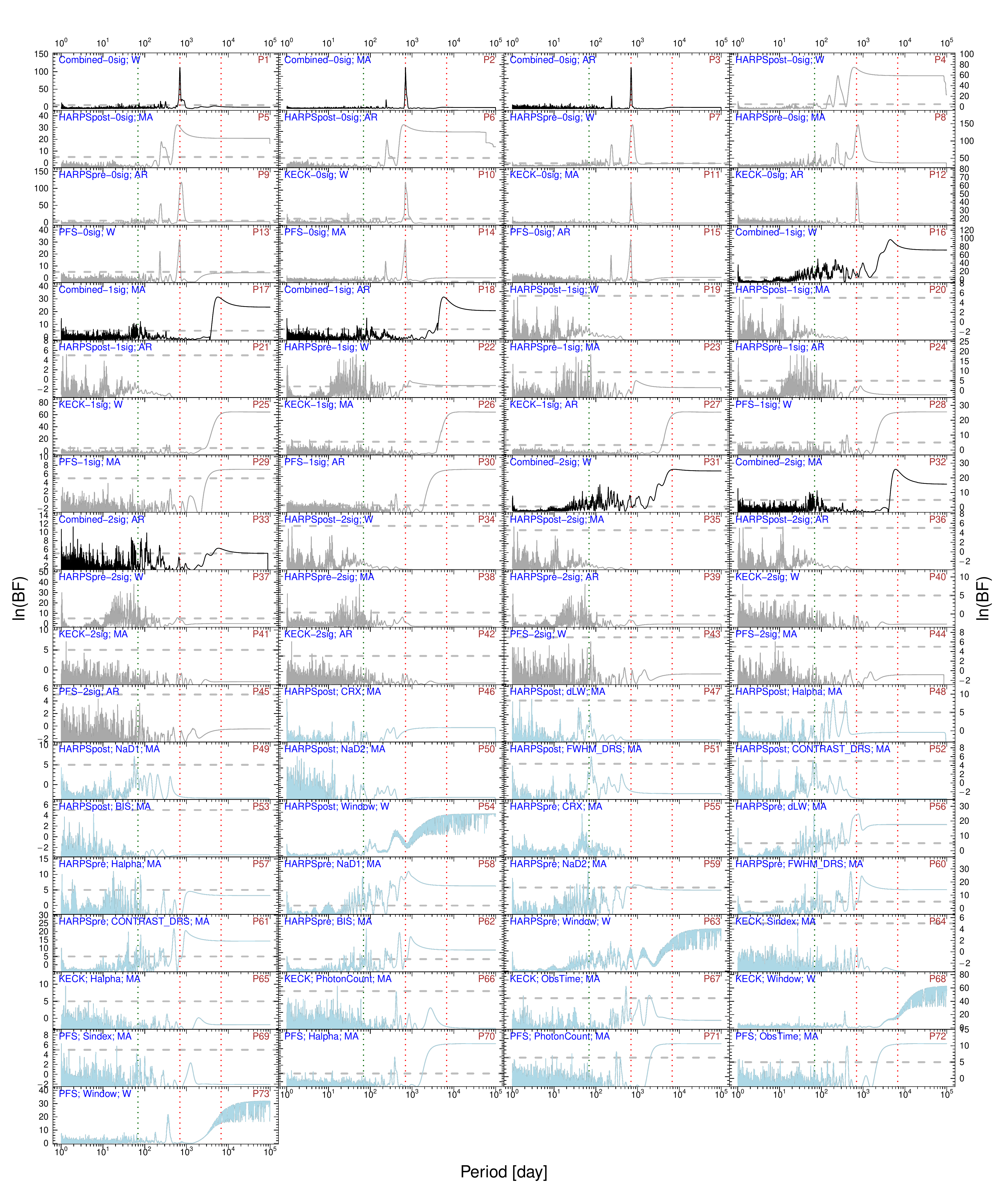}
  \caption{BFPs for the RVs and noise proxies for GJ 317. The dark
    green dotted lines denote the literature 69\,d rotation period. The
    red lines denote the 696\,d and 6719\,d Keplerian signals. }
  \label{fig:GJ317}
\end{figure*}

\begin{figure*}
  \centering
  \includegraphics[scale=0.5]{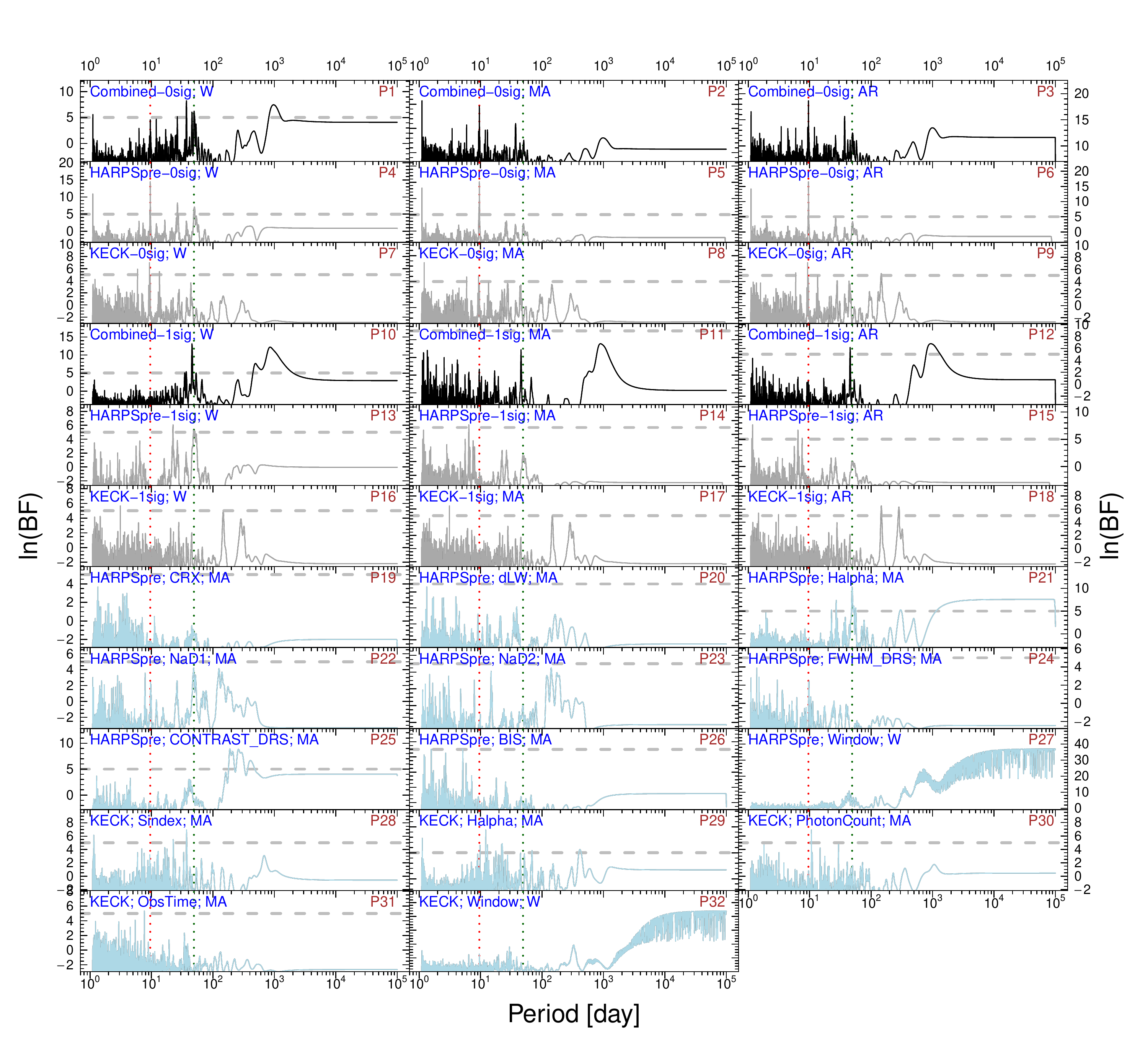}
  \caption{BFPs for the RVs and noise proxies for GJ 480. The dark
    green dotted lines denote the 49.3\,d rotation period determined
    in this work. The red lines denote the 9.56\,d Keplerian signal. }
  \label{fig:GJ480}
\end{figure*}

\begin{figure*}
  \centering
  \includegraphics[scale=0.5]{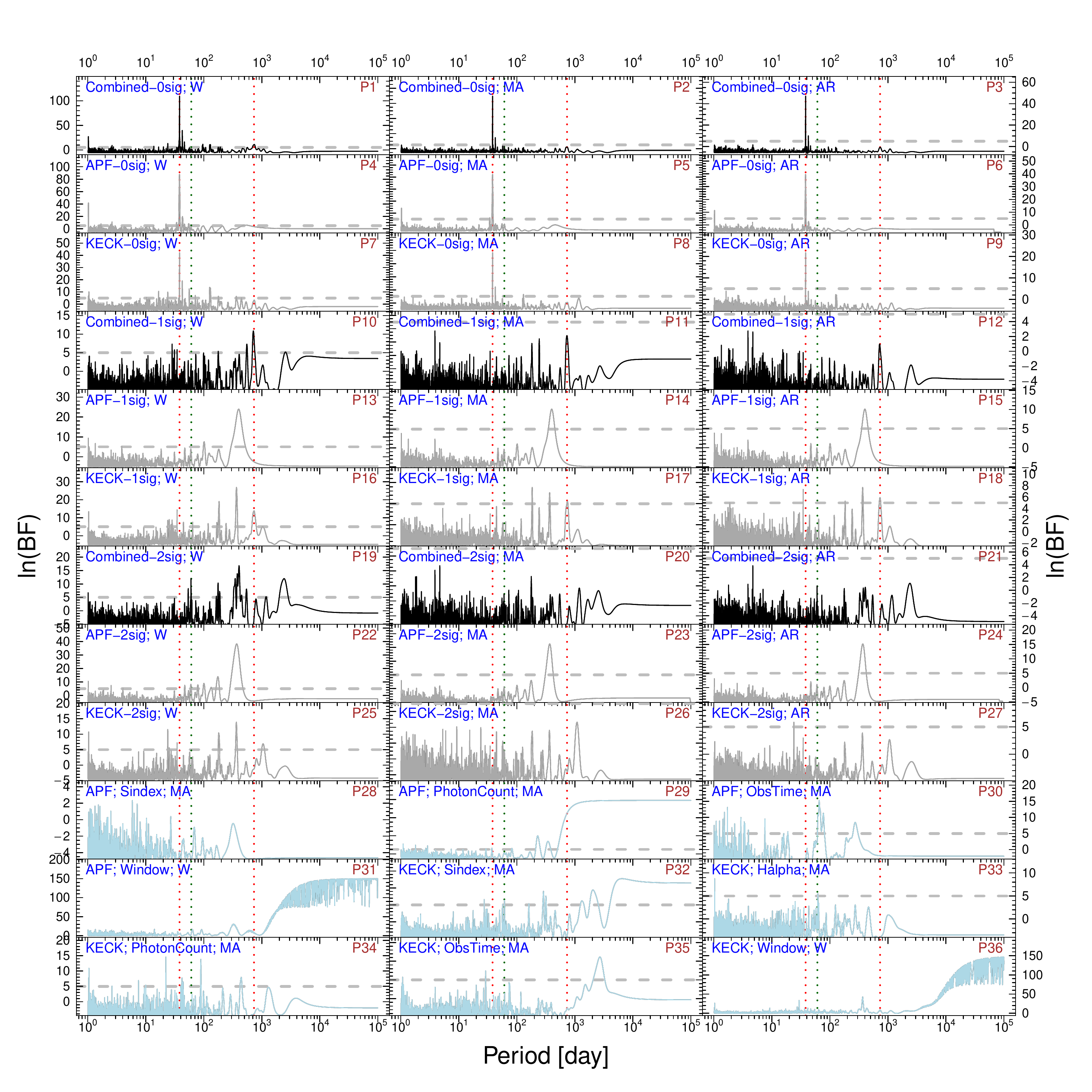}
  \caption{BFPs for the RVs and noise proxies for GJ 687. The dark
    green dotted lines denote the literature 60.8\,d rotation period. The
    red lines denote the 38.1\,d and 726\,d Keplerian signals. }
  \label{fig:GJ687}
\end{figure*}

\begin{figure*}
  \centering
  \includegraphics[scale=0.5]{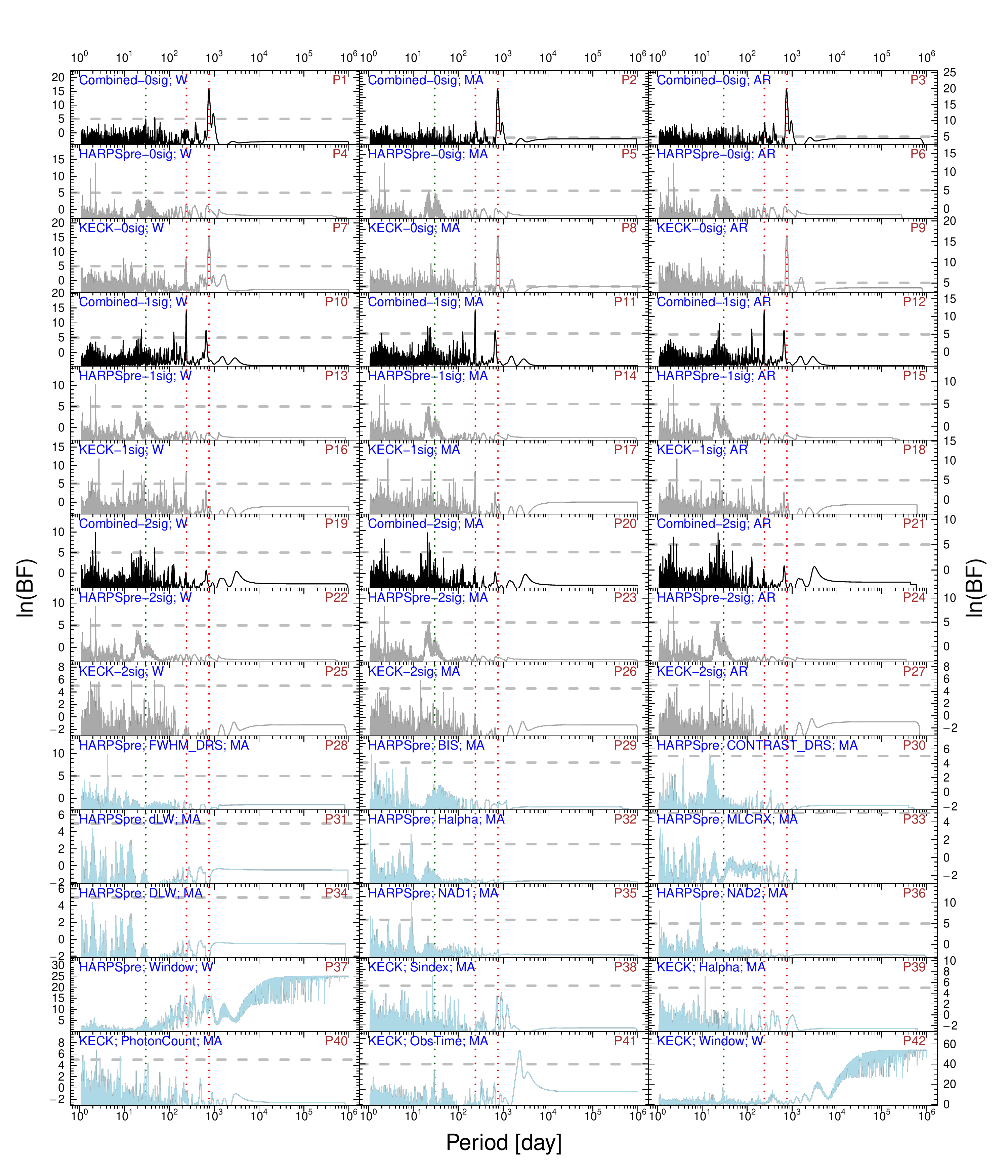}
  \caption{BFPs for the RVs and noise proxies for GJ 9066. The dark
    green dotted lines denote the literature 30\,d rotation period. The
    red lines denote the 242\,d and 773\,d Keplerian signals. }
  \label{fig:GJ9066}
\end{figure*}

\begin{figure*}
  \centering
  \includegraphics[scale=0.5]{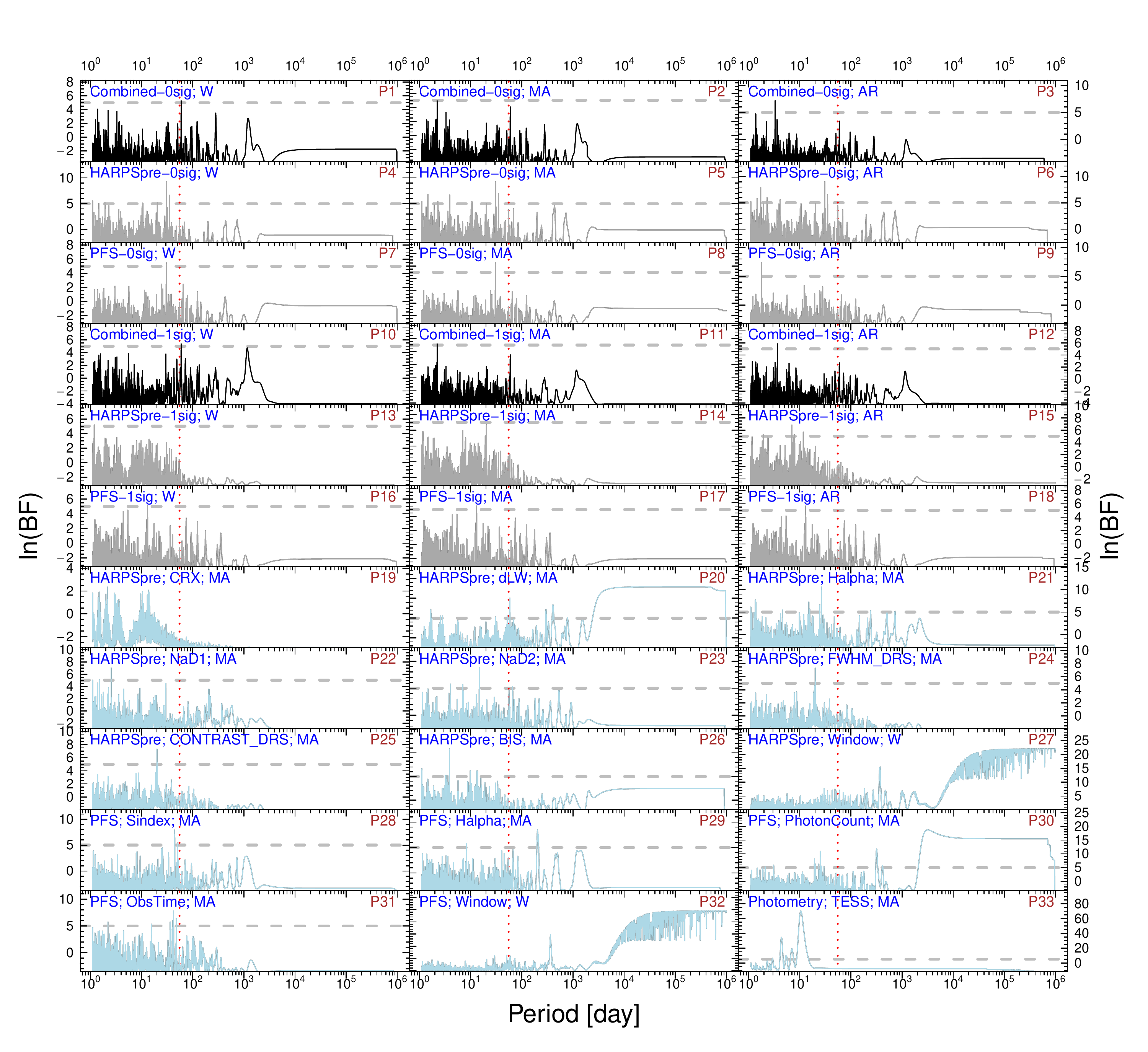}
  \caption{BFPs for the RVs and noise proxies for HIP 107772. The
    red lines denote the 55.3\,d Keplerian signal. }
  \label{fig:HIP107772}
\end{figure*}

\begin{figure*}
  \centering
  \includegraphics[scale=0.5]{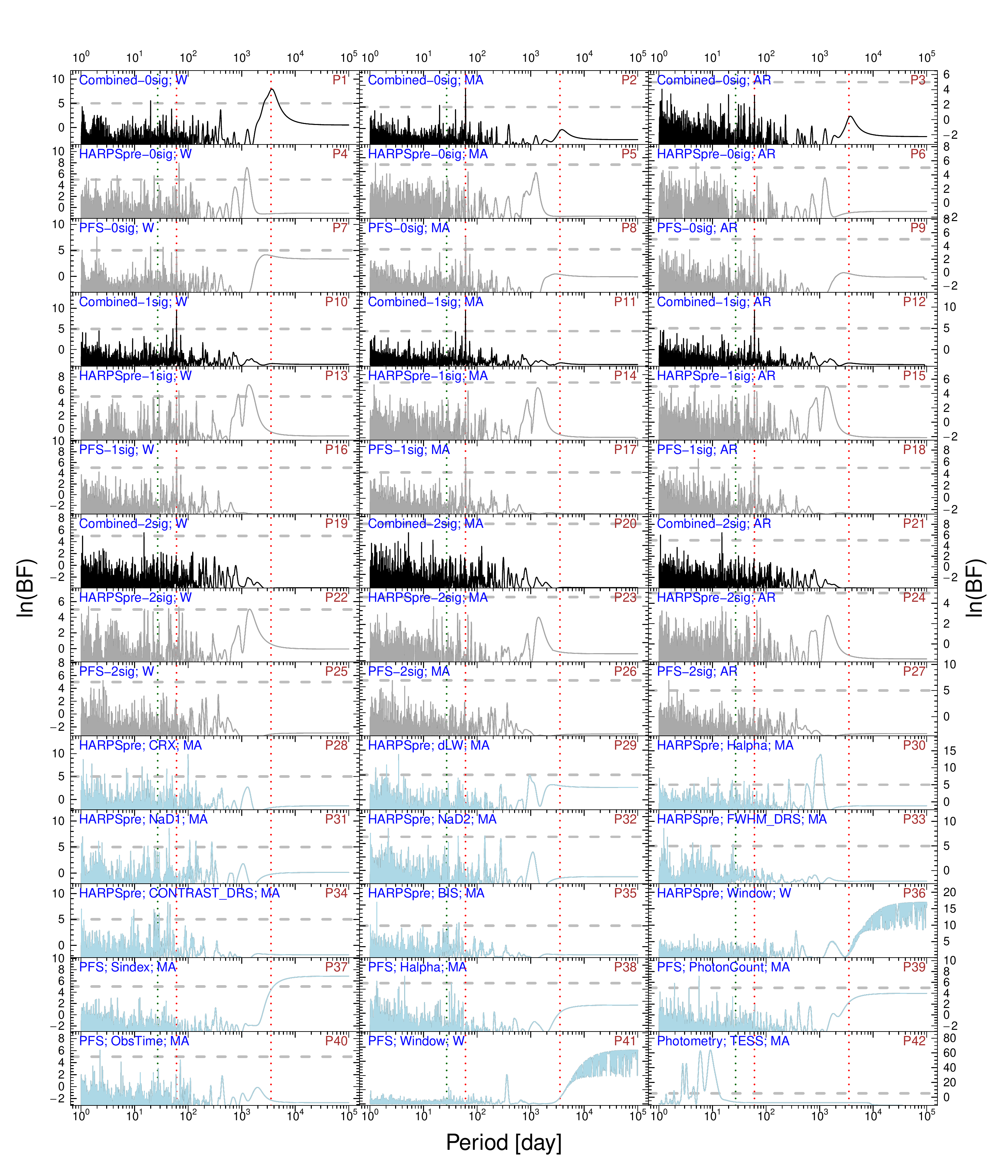}
  \caption{BFPs for the RVs and noise proxies for HIP 38594. The dark
    green dotted lines denote the literature 27\,d rotation period. The
    red lines denote the 60.7\,d and 3525\,d Keplerian signals. }
  \label{fig:HIP38594}
\end{figure*}

\begin{figure*}
  \centering
  \includegraphics[scale=0.5]{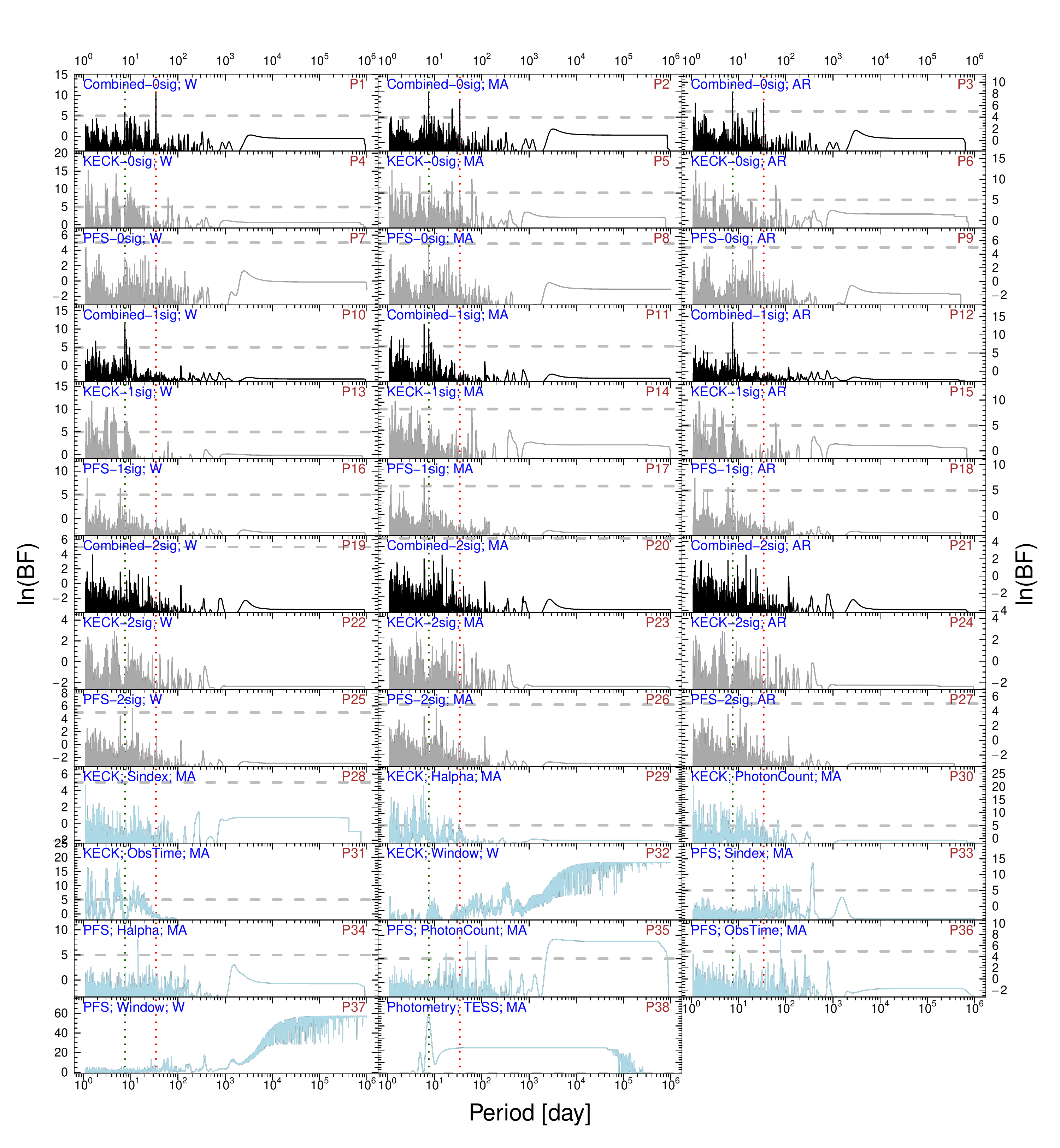}
  \caption{BFPs for the RVs and noise proxies for HIP4845. The dark
    green dotted lines denote the 7.6\,d rotation period determined in
    this work. The red lines denote the 34.2\,d Keplerian signal. }
  \label{fig:HIP4845}
\end{figure*}

\begin{figure*}
  \centering
  \includegraphics[scale=0.5]{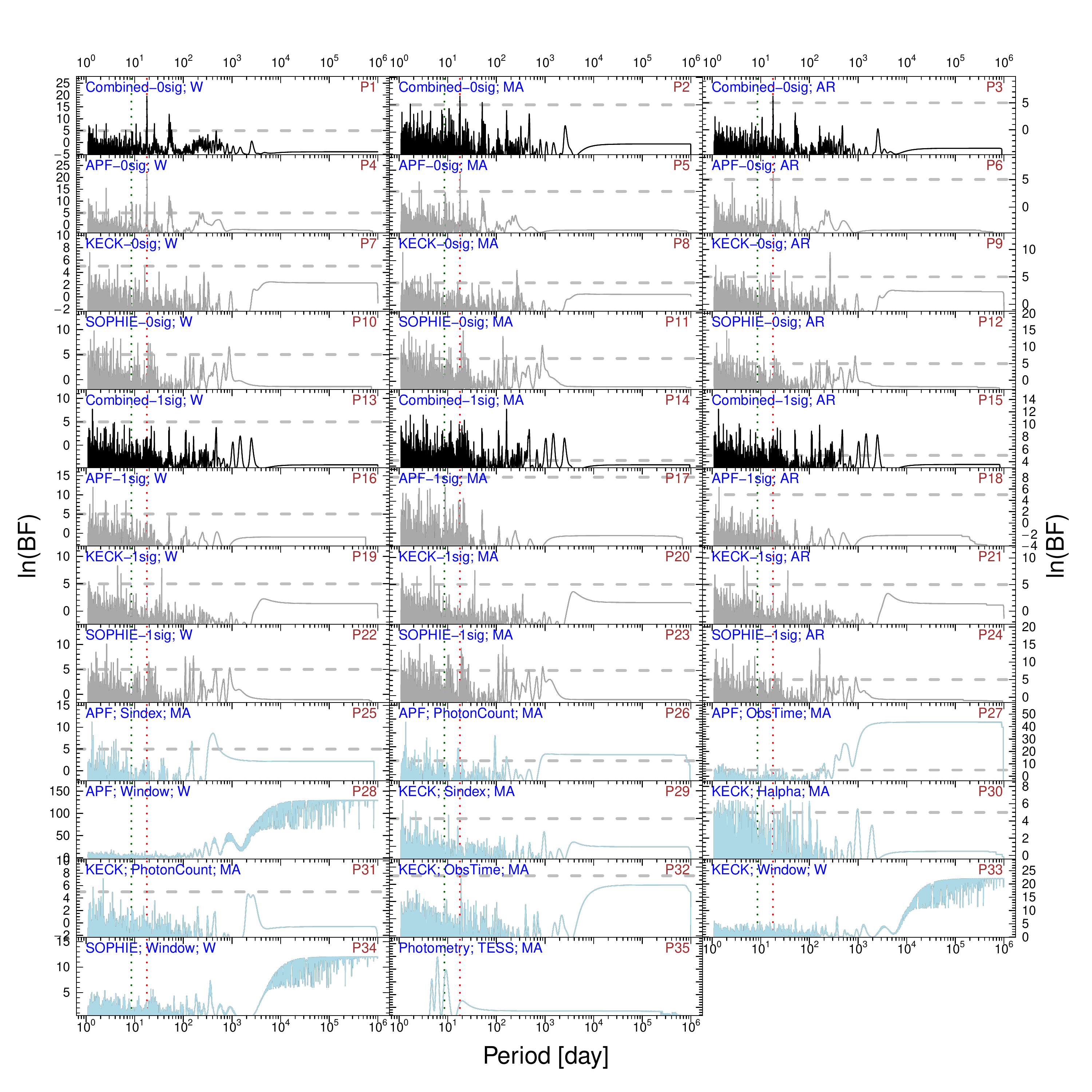}
  \caption{BFPs for the RVs and noise proxies for HIP 48714. The dark
    green dotted lines denote the literature 8.55\,d rotation period. The
    red lines denote the 17.8\,d Keplerian signal. }
  \label{fig:HIP48714}
\end{figure*}

\begin{figure*}
  \centering
  \includegraphics[scale=0.5]{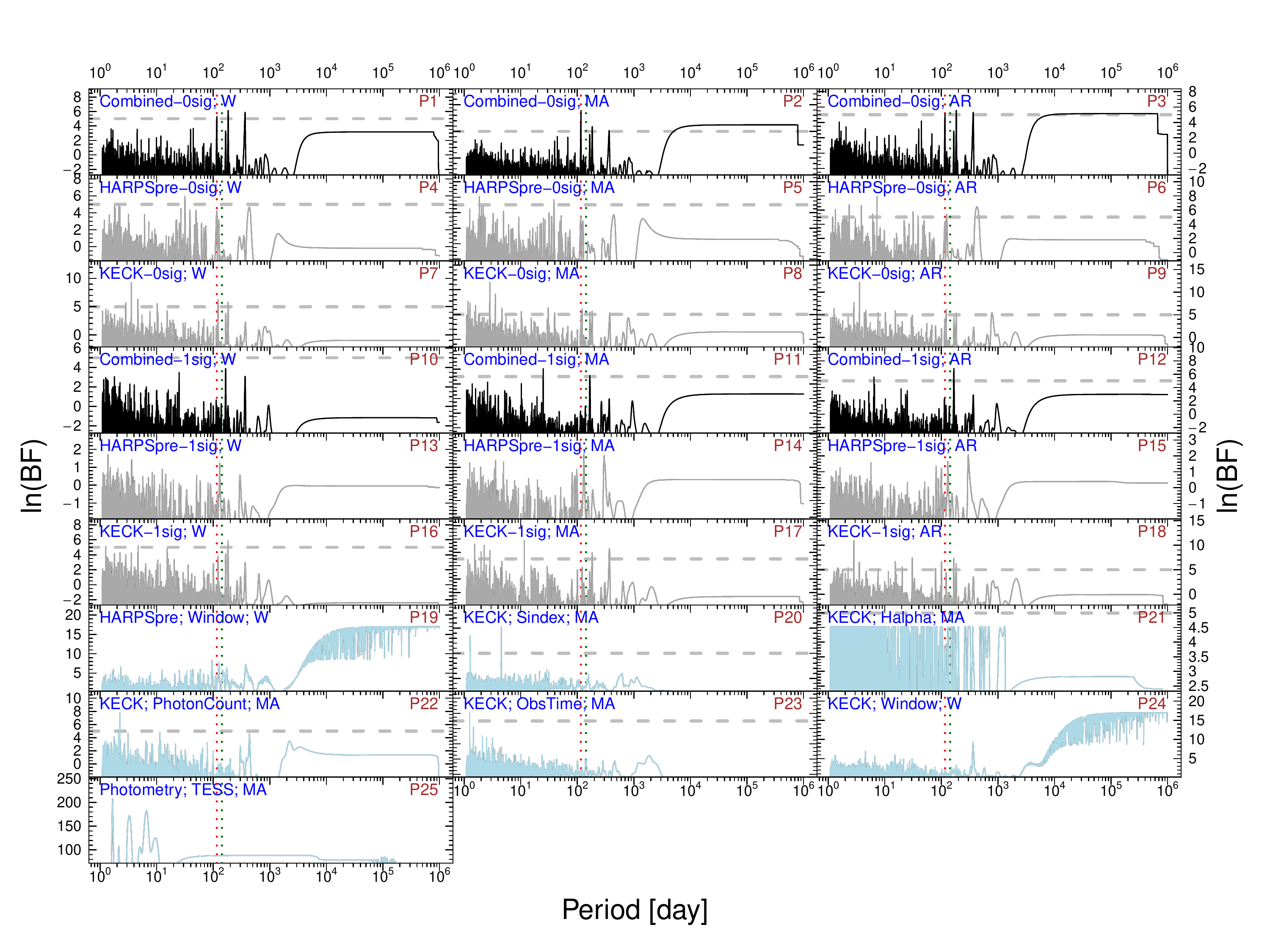}
  \caption{BFPs for the RVs and noise proxies for HIP 60559. The dark
    green dotted lines denote the literature 143\,d rotation period. The
    red lines denote the 116\,d Keplerian signal. }
  \label{fig:HIP60559}
\end{figure*}

\begin{figure*}
  \centering
  \includegraphics[scale=0.5]{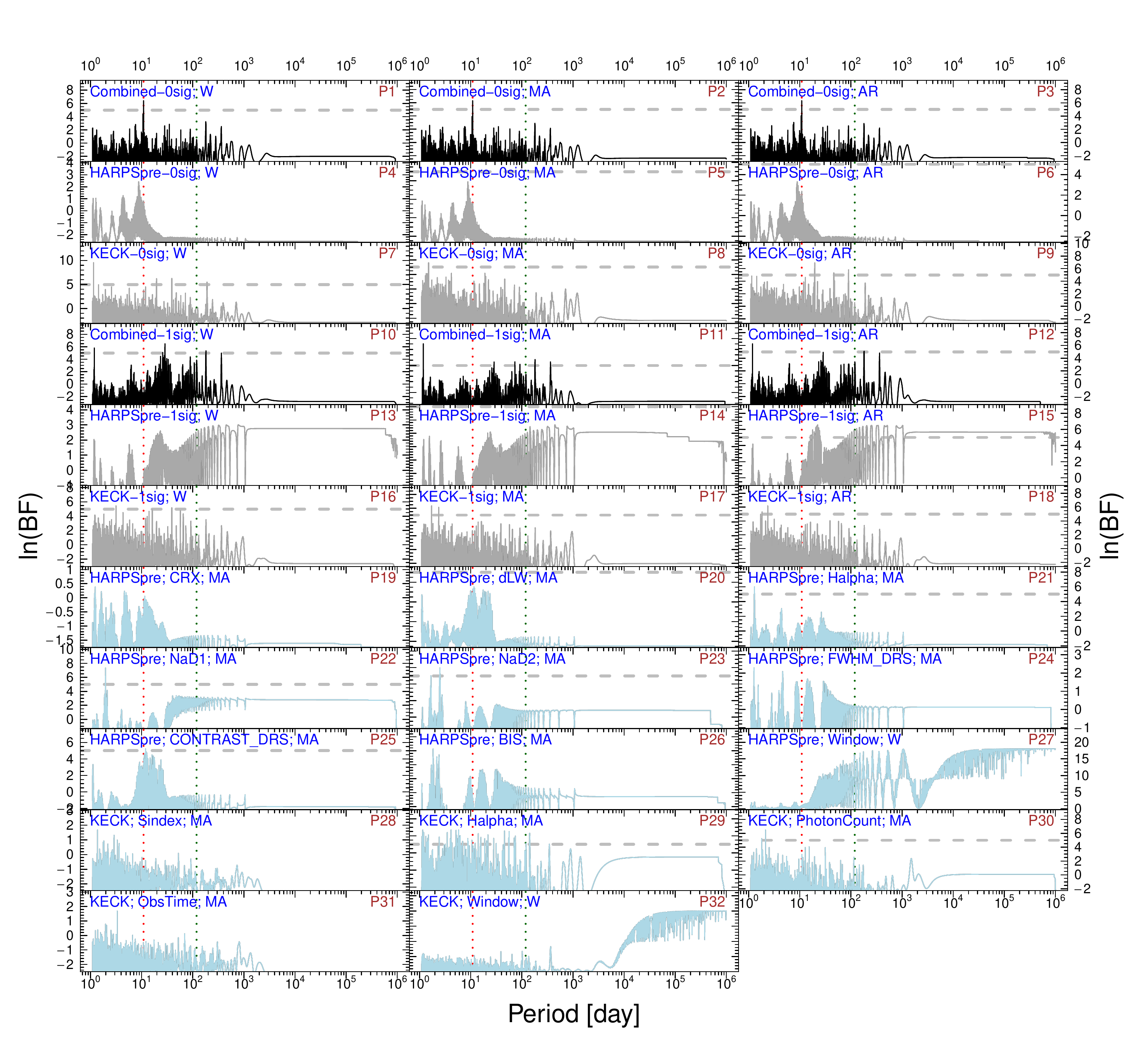}
  \caption{BFPs for the RVs and noise proxies for HIP 67164. The dark
    green dotted lines denote the literature 119\,d rotation period. The
    red lines denote the 10.9\,d Keplerian signal. }
  \label{fig:HIP67164}
\end{figure*}

\begin{itemize}
\item{\bf GJ 2056} (HIP 34785) is an M0-type star with a rotation
  period of about 32\,days \citep{astudillo17} based on a study of CaII H and K emission lines. As is seen in Fig. \ref{fig:GJ2056}, the
  signal at a period of 3000 days can be identified in the BFPs for
  different noise models (P1 to P3) although the phase is not well covered by the
  data, as shown in Fig. \ref{fig:phase}. The 69.9 day signal corresponds to a
  planet candidate on an eccentric orbit, leading to a low power in
  the BFPs that assume zero eccentricity. On the other hand, the
  69.9 day signal is unique in the MP (Fig. \ref{fig:MP}) despite low
  power in the early epochs that are sparsely sampled. The MP is not
  as useful for the 3000 day signal because its period is comparable
  with the observational baseline. Considering that the orbital
    phase, especially the periastron, of this candidate is not well
    sampled by the RVs (see Fig. \ref{fig:phase}), we regard it as a
    weak Neptune candidate. We confirm GJ 2056 b as a strong Neptune candidate located in
  the HZ and its potential moons might be habitable. 
  
\item{\bf GJ 317} (LHS 2037) is an M dwarf with a rotation period of
  about 69 days according to \citep{astudillo17}. Two signals at periods
  of 692 and $>$7100 days have been identified by \cite{anglada12b}
  using KECK data in combination with astrometric data. With HARPS, KECK and PFS data, we are able to constrain the orbit of GJ 317 b better and identify GJ 317 c as a Jupiter analog with an orbital period of 6700 days and a minimum mass of $2.13\pm 0.19$\,$M_{\rm Jup}$. Its semi-major axis is 5.9$\pm$0.28\,au, similar to the distance from Jupiter to the Sun (5.20\,au). Since GJ 317 is only about 15\,pc from us, GJ 317 c is about 0.4$''$ from GJ 317 and thus is
  detectable by the CGI of WFIRST \citep{tang19}. It is evident from
  the fit to combined set in Fig. \ref{fig:phase} that these two signals
  are very significant. This is also demonstrated in the MP shown in
  Fig. \ref{fig:MP} and the BFPs in Fig. \ref{fig:GJ317}.

\item{\bf GJ 480} (Wolf 433 or HIP 61706) is a high proper motion red
  dwarf. We confirm the detection of this signal with a comprehensive analysis of the combined HARPS and KECK data. The
  signal is robust to the choice of noise models (P1-P3 in
  Fig. \ref{fig:GJ480}) and is consistent over time
  (Fig. \ref{fig:MP}). We also identify an activity signal at a
  period of $49.3\pm 0.2$\,d, which is significant in the BFPs for
  NaD1 (P22 in Fig. \ref{fig:GJ480}) and H$\alpha$ (P21) of the HARPSpre data. 

\item{\bf GJ 687} (LHS 450 or HIP 86162) is a red dwarf with a
  rotation period of about 60 days \citep{burt14}. It is found to host at least one
  planet at a period of 38.14\,days \citep{burt14}. In our combined analysis of APF, KECK, and SOPHIE,
  we confirm previous findings and improve the parameter
  estimation. However, we find a solution with higher eccentricity for
  the 758 day signal probably due to the broader Gaussian prior adopted
  for eccentricity in this work. This signal is not sensitive to the
  choice of noise models (P1-P3 in Fig. \ref{fig:GJ687}). The 38.1\,d
  signal is found in both the APF and KECK sets (P4-P9) while the 726\,d
  signal is only found in the KECK set because the APF baseline is too
  short for such a long period signal (see the raw data in
  Fig. \ref{fig:MP}). Hence we consider these two signals as strong
  planet candidates.
  
\item{\bf GJ 9066} (LHS 11 or GJ 83.1) is an eruptive variable red
  dwarf with a rotation period of about 30\,days
  \citep{astudillo17}. Based on our combined analysis of the HARPS and KECK data, we regard the signals at periods of 769
  and 242 days as robust planet candidates. The 769 and 242
  day signals are significant in the KECK data (P7-P9 and P16-P18 in
  Fig. \ref{fig:GJ9066}). The 30-day rotation signal is significant in
  the BFP for KECK S-index (P38). The 242 day signal is quite
  consistent over time while the 773 day signal is more significant in
  recent epochs than in previous ones due to recent high cadence
  sampling (Fig. \ref{fig:MP}). In particular, these two planet
  candidates form a 3:1 mean motion resonance, which may stabilize the
  system over long timescales. 

\item{\bf HIP 107772} (TYC 7986-911-1) is a red dwarf without any
  known planets. A signal around 55 days is found to be
  significant. This signal fit the RV data well (see
  Fig. \ref{fig:phase}) and it is consistently significant over time (Fig. \ref{fig:MP}). This signal is identifiable in the BFPs for
  different noise models (P1-P3 in Fig. \ref{fig:HIP107772}) and
  different data sets (P2-P9 in Fig. \ref{fig:HIP107772}). It does not
  overlap with the activity signals (P19-P33 in
  Fig. \ref{fig:HIP107772}). Hence this signal corresponds to a strong
  Neptune candidate located in the HZ. 

\item{\bf HIP 38594} (Ross 429) is a red dwarf rotating with a period
  of about 27\,days \citep{astudillo17}. Two signals at periods of
  60.7 and 3480 days are found to be significant based on the combined
  analysis of the HARPS and PFS data. The MP shows good time
  consistency for HIP 38594 b despite cadence dependent variation in
  power (Fig. \ref{fig:MP}). However, the period of HIP 38594 c is
  too long for time consistency test although the signal is apparent
  in the residual RVs (panel for HIP 38594 c in
  Fig. \ref{fig:MP}). In particular, HIP 38594 b is a super-Earth
  located in the optimistic HZ, as shown in Fig. \ref{fig:hz}. We
  regard the two signals as strong planet candidates.

\item{\bf HIP 4845} (GJ 3072) is an M dwarf without known
  planet. Through combined analysis of the HARPS, KECK, and PFS data,
  we find two signals at periods of 7.6 and 34.2\,days. The former is
  found to be significant in the periodogram for the TESS data (P38 of
  Fig. \ref{fig:HIP4845}) while the later corresponds to a warm
  super-Earth. The MP for the 34.2 day signal (Fig. \ref{fig:MP}) shows
  consistent significance over time. This signal is
  less significant in recent epochs because of low cadence
  sampling. It is robust to the choice of noise models (P1-P3 and
  P10-P12 in Fig. \ref{fig:HIP4845}) and is identifiable in the KECK
  and PFS individual sets (e.g., P5, P8, P14, and P17 in
  Fig. \ref{fig:HIP4845}). Thus we regard the 34.2 day signal as a
  strong planet candidate. Although the 7.6 day signal is as
  significant and consistent as the 34.2 day signal, it overlaps with
  the signal found in TESS photometric data, suggesting an activity
  origin. This demonstrates the importance of a comprehensive
  diagnostics of activity signals, which are sometimes very similar to
  Keplerian signals. 

\item{\bf HIP 48714} (GJ 373 or LHS 2211) is a red dwarf with a
  rotation period of 8.55\,days \citep{oelkers18}. A super-Earth with
  an orbital period of 17.8\,days is found to orbit around the star
  based on our combined analysis of the APF, KECK, and SOPHIE data. As
  shown in the MP in Fig. \ref{fig:MP}, the 17.8 day signal is
  especially significant in recent epochs dominated by high cadence
  APF data. Since the earlier epochs are not well sampled, such
  inconsistency is not due to the intrinsic time variability of the
  signal. As shown in Fig. \ref{fig:HIP48714}, the signal is robust to
  the choice of noise models (P1-P3). It is significant in the APF set (P4)
  and is identifiable in the KECK set (P7). The signal does not
  overlap with activity signals (P25-P35). Therefore we conclude that
  the 17.8 day signal is a strong super-Earth candidate. 

\item{\bf HIP 60559} (Ross 695) is a red dwarf with a rotation period
  of about 143\,days \citep{astudillo17}. A signal at a period of
  116\,days is identified based on the combined analysis of HARPS and
  KECK data. The signal is not sensitive to the choice of noise models
  (P1-P3 in Fig. \ref{fig:HIP60559}) and is identifiable in both the HARPS
  and KECK data sets. As shown in Fig. \ref{fig:MP}, the period of
  the 116 day signal seems to vary slightly due to aliasing and low
  cadence sampling. Thus we consider it as a weak Neptune candidate.
  
\item{\bf HIP 67164} (LHS 2794 or GJ 3804) is a red dwarf with a
  rotation period of about 119\,days \citep{astudillo17}. A signal at
  a period of 10.9\,days is identified through the combined analysis of
  HARPS and KECK data. The signal is robust to the choice of
  noise models (P1-P3 in Fig. \ref{fig:HIP67164}) and is identifiable
  in the HARPS (P6) and KECK sets (P7). However, due to the highly
  irregular sampling of the data, it does not show consistent significance over time in
  the MP. We consider this signal as a weak candidate. 
\end{itemize}

\section{Dynamical stability}\label{sec:stability}
In the same manner as paper II, we examine the dynamical stability of the new planet candidates with a large suite of numerical simulations utilizing the $Mercury6$ mixed-variable symplectic integrator \citep[MVS;][]{chambers99}.  These simulations are designed to quickly identify unstable regions of parameter space within our calculated uncertainties for the planets' orbital elements (Table \ref{tab:planet}). Thus, while a definitive proof of each systems' stability is beyond the scope of this manuscript, systems that evolve regularly in each of our various realizations are highly likely to be stable (though we do not consider possible perturbations from additional, undetected planets).

\begin{figure}
	\centering
	\includegraphics[width=.45\textwidth]{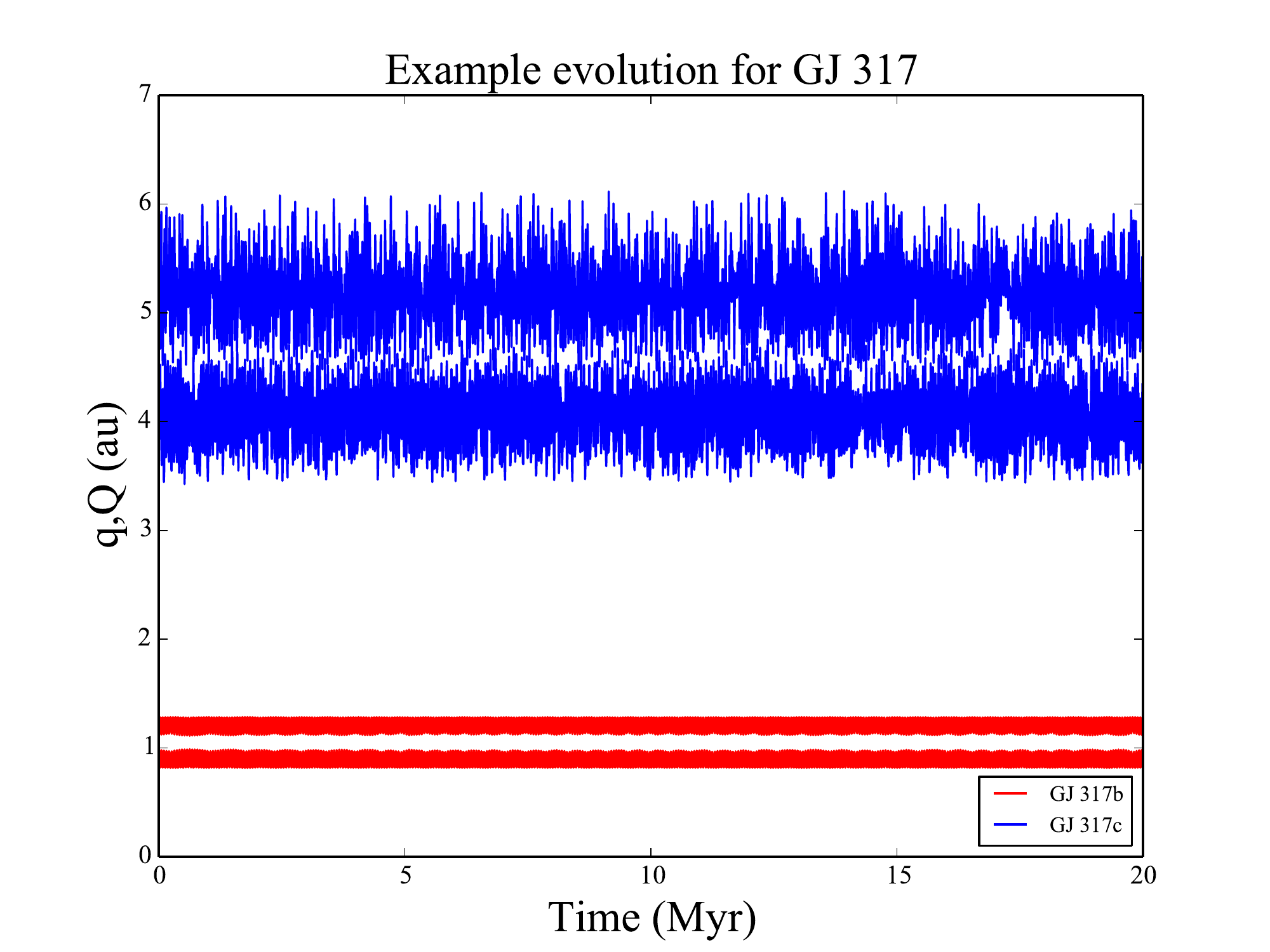}
        \includegraphics[width=.45\textwidth]{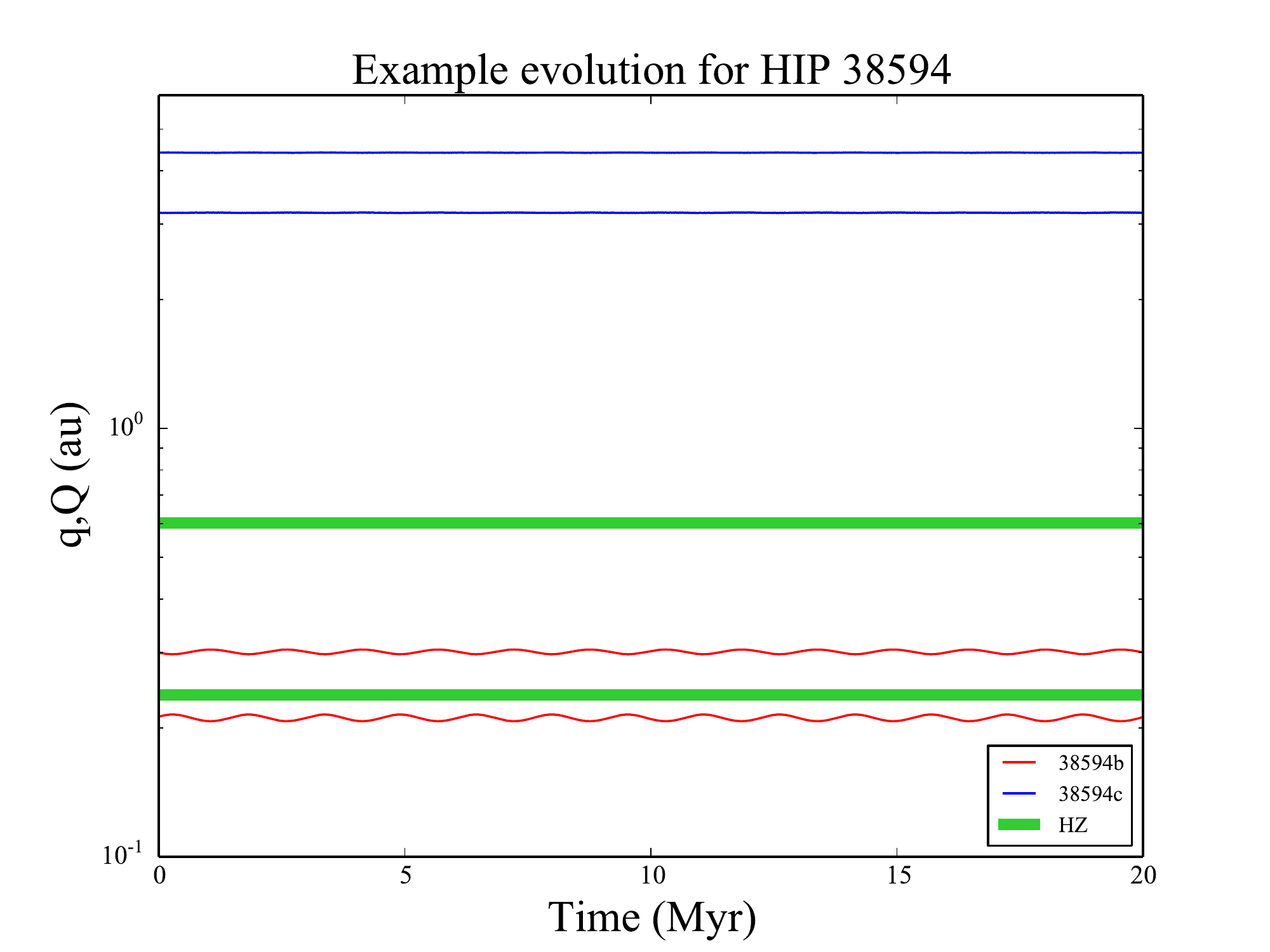}
	\caption{Example evolutionary scheme from two of our dynamical
          simulations studying the GJ 317 and HIP 38594 systems. The
          pericenter and apocenter for each planet are respectively
          plotted with blue (GJ 317 c and HIP 38594 c) and red (GJ 317
          b and HIP 38594 b) lines. The inner and outer edges of the
          HZ of HIP 38594 are shown by horizontal green lines.}
	\label{fig:qaq}
\end{figure}

For each multi-planet system we consider a grid of five eccentricities
and masses for each object within the ranges of uncertainties for the
respective parameters reported in Table \ref{tab:planet}. As in paper II, we also analyze three possible orientations for each planetary system: $I=$ 30, 60 and 90$^{\circ}$. Thus, each individual system is scrutinized with 1,875 separate numerical simulations. Planetary inclinations are selected randomly from nearly co-planar distributions, and the remaining angular orbital elements (i.e.: those not listed in Table \ref{tab:planet}) are selected at random from uniform distributions. Each system is integrated for 1 Myr utilizing a time-step equal to $\sim$5$\%$ of the inner planet's orbital period \citep[e.g.:][]{gilbert20}. Systems containing at least one planet with $e>$0.5 are integrated for 20\,Myr to account for high-eccentricity dynamics.  The results of our dynamical analysis are summarized as follows:

\begin{itemize}
	\item \textbf{Systems exhibiting regular behavior.} Within
          our tested parameter space, the two planet systems GJ 2056,
          GJ 317, GJ 9066, and HIP 38594 display no evidence of
          instability or chaotic evolution \citep[e.g.:][]{laskar97}.
          In all cases, the planets' orbits are governed by regular
          secular oscillations in $e$; the magnitude of which are
          related to their masses and initial eccentricities
          \citep[e.g.:][]{dermott99}.  The largest such oscillations
          occur in GJ 317, the system possessing the most massive
          planets.  An example of the evolution of this system is
          plotted in the left panel of Fig. \ref{fig:qaq}.  As GJ 317
          b and c are well separated in terms of their orbital period
          ratio for all of our tested combinations of semi-major axes
          ($P_{c}/P_{b}\sim$ 10), our simulations suggest that this
          system is dynamically stable.

        \item \textbf{Dynamical stability of HIP 38594.} We also study the dynamical
          stability of HIP 38594, the host of an HZ super-Earth. As
          shown in the right panel of Fig. \ref{fig:qaq}, HIP 38594 b
          migrates into and out of the HZ when approaching its
          apocenter and pericenter, respectively. However, such an
          non-circular orbit might not be representative because the eccentricity
          given in Table \ref{tab:planet} is consistent with zero at
          the 2-sigma confidence level. Assuming a circular orbit,
          HIP 38594 b would be stable over at least a few millions in
          the HZ. 

        \item \textbf{System with unstable parameter space: GJ 687.}  In 5$\%$ of our simulations investigating the stability of GJ 687, the inner planets' pericenter was excited to the point that the planet collided with the central body.  This occurred exclusively in our integrations testing the largest eccentricities ($\simeq$0.62) and masses ($M\gtrsim$ 36 $M_{\oplus}$ for the I$=$30$^{\circ}$ case) for the outer planet.  In these isolated instances, the planets begin on nearly-crossing orbits where they interact strongly with each other.  The more massive outer planet's eccentric forcing on GJ 687 b drives large secular oscillations in the smaller planet's eccentricity, eventually driving its pericenter on to a collision course with the central star.  As the average timescale for the loss of GJ 687 b in our simulations ($\sim$300 Kyr) is significantly less than the system's age, and the planets evolve regularly within the remainder of our tested parameter space, we conclude that the system is indeed stable.  Thus, our results imply additional constraints on the eccentricity and mass of GJ 687 c, likely limiting them to the lower range of the values reported in table 2 (specifically, $e\lesssim$ 0.51 for nominal mass values and $e\lesssim$ 0.40 for $M_{c}\gtrsim$ 36 $M_{\oplus}$).
\end{itemize} 

\section{Discussion and conclusion}\label{sec:conclusion}
In this work, we identify ten strong planet candidates as well as
three weak candidates, and confirm three previous candidates. Weak candidates
need followup investigations to confirm. The strong planet
candidates satisfy the planet selection criteria and are unlikely to be
caused by stellar activity based on our diagnostics. We also confirm previous candidates and
improve their orbital solutions through our independent
analyses. Among these planet candidates, there are one temperate
super-Earth, four hot super-Earths, three temperate Neptunes, four
cold Neptunes, and four cold Jupiters. 

To date, HIP 38594 is the most massive M dwarf host of temperate Earths and super-Earths
that are found through the RV method. As an early-type M dwarf, HIP 38594
is a Goldilocks M dwarf host for habitable planets because it is less
active and has larger habitable zone than other types of M dwarfs. We also
investigate the dynamical stability of the HZ planet, HIP 38594 b, and find that the orbit of the planet partially overlaps with the
HZ. Considering that the orbital solution for this planet is
consistent with a circular orbit, HIP 38594 b is probably on a nearly
circular orbit and is thus unlikely to migrate out of the HZ
frequently.


We also detect three temperate Neptunes and four cold Neptunes, contributing
significantly to a rarely explored population. On the other hand, we
find four cold Jupiters, equal to the number of cold Neptunes. Considering
the fact that cold Jupiters induce larger RV semi-amplitudes and are
thus easier to detect, the sample of cold Neptunes and Jupiters
detected in this work are consistent with a high occurrance rate of
cold Neptunes inferred from microlensing observations by \cite{suzuki16}. They conclude that ``cold Neptunes
are likely to be the most common type of planets beyond the snow
line.'' Cold Neptunes have rarely been detected through the RV method until the
recent accumulation of large amount of RV data. Due to their
relatively large angular separation from their hosts, the cold
Neptunes and cold Jupiters detected in this work are good targets for direct
imaging by future facilities such as HabEx and ELT. 

Our discovery of multiple planets in RV data demonstrates the
feasibility of a comprehensive RV survey of nearby planets, especially
Earth analogs. Based on combined analyses of all available RV data for
M dwarfs, we are able to select the most promising targets for further
analyses and followup obsevations. This leads to the identification of
smaller planets embedded in noisy RV time series obtained by
different groups. Our stellar activity diagnostics allow us to classify
signals into different categories by accounting for their consistency
over time, robustness to the choice of noise models, and overlaps with
activity signals. In particular, the MPs visualize the time consistency of signals, the
BFPs test the sensitivity of signals to noise models and stellar
activity. Nevertheless, stellar activity is still the major challenge
to the detection of smaller signals caused by Earth twins. According
to our analyses of the RV data for nearby M dwarfs in paper II and
this paper, stellar variability sets the current limit of detectable RV signals for M dwarfs to be $\sim$1\ms. Our detection of
temperate super-Earths and cold Neptunes around early-type M dwarfs slightly above this limit suggest a large undetected population of small planets embedded in the current RV data.

\section*{Acknowledgements}
This work has made use of data from the European Space Agency (ESA)
mission {\it Gaia} (https://www.cosmos.esa.int/gaia), processed by the
{\it Gaia} Data Processing and Analysis Consortium (DPAC,
https://www.cosmos.esa.int/web/gaia/dpac/consortium). Funding for the
DPAC has been provided by national institutions, in particular the
institutions participating in the {\it Gaia} Multilateral
Agreement. This research has also made use of the Keck Observatory Archive
(KOA), which is operated by the W. M. Keck Observatory and the
NASA Exoplanet Science Institute (NExScI), under contract with
the National Aeronautics and Space Administration.  This
research has also made use of the services of the ESO Science
Archive Facility, NASA's Astrophysics Data System Bibliographic
Service, and the SIMBAD database, operated at CDS, Strasbourg, France.
Support for this work was provided by NASA through Hubble
Fellowship grant HST-HF2-51399.001 awarded by the Space Telescope
Science Institute, which is operated by the Association of
Universities for Research in Astronomy, Inc., for NASA, under contract
NAS5-26555. The authors acknowledge the years of technical support
from LCO staff in the successful operation of PFS, enabling the
collection of the data presented in this paper. We would also like to
acknowledge the many years of technical support from the UCO/Lick
staff for the commissioning and operation of the APF facility atop
Mt. Hamilton. Part of this research was carried out at the Jet Propulsion Laboratory, California Institute of Technology, under a contract with the National Aeronautics and Space Administration (NASA).
\software{R package magicaxis \citep{robotham16},
fields \citep{fields}, MASS \citep{ripley13}, minpack.lm \citep{elzhov16}.}

\bibliographystyle{aasjournal}
\bibliography{nm}  

\begin{thebibliography}{}
\expandafter\ifx\csname natexlab\endcsname\relax\def\natexlab#1{#1}\fi
\providecommand{\url}[1]{\href{#1}{#1}}
\providecommand{\dodoi}[1]{doi:~\href{http://doi.org/#1}{\nolinkurl{#1}}}
\providecommand{\doeprint}[1]{\href{http://ascl.net/#1}{\nolinkurl{http://ascl.net/#1}}}
\providecommand{\doarXiv}[1]{\href{https://arxiv.org/abs/#1}{\nolinkurl{https://arxiv.org/abs/#1}}}

\bibitem[{{Anglada-Escud{\'e}} {et~al.}(2012){Anglada-Escud{\'e}}, {Boss},
  {Weinberger}, {Thompson}, {Butler}, {Vogt}, \& {Rivera}}]{anglada12b}
{Anglada-Escud{\'e}}, G., {Boss}, A.~P., {Weinberger}, A.~J., {et~al.} 2012,
  \apj, 746, 37, \dodoi{10.1088/0004-637X/746/1/37}

\bibitem[{Anglada-Escud{\'e} {et~al.}(2016)Anglada-Escud{\'e}, Amado, Barnes,
  Berdinas, Butler, Coleman, de~la Cueva, Dreizler, Endl, Giesers,
  {et~al.}}]{anglada16}
Anglada-Escud{\'e}, G., Amado, P.~J., Barnes, J., {et~al.} 2016, Nature, 536,
  437

\bibitem[{{Artigau} {et~al.}(2014){Artigau}, {Kouach}, {Donati}, {Doyon},
  {Delfosse}, {Baratchart}, {Lacombe}, {Moutou}, {Rabou}, {Par{\`e}s},
  {Micheau}, {Thibault}, {Reshetov}, {Dubois}, {Hernandez}, {Vall{\'e}e},
  {Wang}, {Dolon}, {Pepe}, {Bouchy}, {Striebig}, {H{\'e}nault}, {Loop},
  {Saddlemyer}, {Barrick}, {Vermeulen}, {Dupieux}, {H{\'e}brard}, {Boisse},
  {Martioli}, {Alencar}, {do Nascimento}, \& {Figueira}}]{artigau14}
{Artigau}, {\'E}., {Kouach}, D., {Donati}, J.-F., {et~al.} 2014, in Society of
  Photo-Optical Instrumentation Engineers (SPIE) Conference Series, Vol. 9147,
  Ground-based and Airborne Instrumentation for Astronomy V, 914715

\bibitem[{{Astudillo-Defru} {et~al.}(2017){Astudillo-Defru}, {Delfosse},
  {Bonfils}, {Forveille}, {Lovis}, \& {Rameau}}]{astudillo17}
{Astudillo-Defru}, N., {Delfosse}, X., {Bonfils}, X., {et~al.} 2017, \aap, 600,
  A13, \dodoi{10.1051/0004-6361/201527078}

\bibitem[{Bechter {et~al.}(2018)Bechter, Bechter, Crepp~Jr, King, \&
  Crass}]{bechter18}
Bechter, A.~J., Bechter, E.~B., Crepp~Jr, J.~R., King, D., \& Crass, J. 2018,
  in Ground-based and Airborne Instrumentation for Astronomy VII, Vol. 10702,
  International Society for Optics and Photonics, 107026T

\bibitem[{{Bryan} {et~al.}(2019){Bryan}, {Knutson}, {Lee}, {Fulton}, {Batygin},
  {Ngo}, \& {Meshkat}}]{bryan19}
{Bryan}, M.~L., {Knutson}, H.~A., {Lee}, E.~J., {et~al.} 2019, \aj, 157, 52,
  \dodoi{10.3847/1538-3881/aaf57f}

\bibitem[{{Burt} {et~al.}(2014){Burt}, {Vogt}, {Butler}, {Hanson}, {Meschiari},
  {Rivera}, {Henry}, \& {Laughlin}}]{burt14}
{Burt}, J., {Vogt}, S.~S., {Butler}, R.~P., {et~al.} 2014, \apj, 789, 114,
  \dodoi{10.1088/0004-637X/789/2/114}

\bibitem[{{Butler} {et~al.}(1996){Butler}, {Marcy}, {Williams}, {McCarthy},
  {Dosanjh}, \& {Vogt}}]{butler96}
{Butler}, R.~P., {Marcy}, G.~W., {Williams}, E., {et~al.} 1996, \pasp, 108,
  500, \dodoi{10.1086/133755}

\bibitem[{{Butler} {et~al.}(2006){Butler}, {Wright}, {Marcy}, {Fischer},
  {Vogt}, {Tinney}, {Jones}, {Carter}, {Johnson}, {McCarthy}, \&
  {Penny}}]{butler06}
{Butler}, R.~P., {Wright}, J.~T., {Marcy}, G.~W., {et~al.} 2006, \apj, 646,
  505, \dodoi{10.1086/504701}

\bibitem[{{Butler} {et~al.}(2017){Butler}, {Vogt}, {Laughlin}, {Burt},
  {Rivera}, {Tuomi}, {Teske}, {Arriagada}, {Diaz}, {Holden}, \&
  {Keiser}}]{butler17}
{Butler}, R.~P., {Vogt}, S.~S., {Laughlin}, G., {et~al.} 2017, \aj, 153, 208,
  \dodoi{10.3847/1538-3881/aa66ca}

\bibitem[{{Chambers}(1999)}]{chambers99}
{Chambers}, J.~E. 1999, \mnras, 304, 793,
  \dodoi{10.1046/j.1365-8711.1999.02379.x}

\bibitem[{{Courcol} {et~al.}(2015){Courcol}, {Bouchy}, {Pepe}, {Santerne},
  {Delfosse}, {Arnold}, {Astudillo-Defru}, {Boisse}, {Bonfils}, {Borgniet},
  {Bourrier}, {Cabrera}, {Deleuil}, {Demangeon}, {D{\'\i}az}, {Ehrenreich},
  {Forveille}, {H{\'e}brard}, {Lagrange}, {Montagnier}, {Moutou}, {Rey},
  {Santos}, {S{\'e}gransan}, {Udry}, \& {Wilson}}]{courcol15}
{Courcol}, B., {Bouchy}, F., {Pepe}, F., {et~al.} 2015, \aap, 581, A38,
  \dodoi{10.1051/0004-6361/201526329}

\bibitem[{{Crane} {et~al.}(2006){Crane}, {Shectman}, \& {Butler}}]{crane06}
{Crane}, J.~D., {Shectman}, S.~A., \& {Butler}, R.~P. 2006, in Society of
  Photo-Optical Instrumentation Engineers (SPIE) Conference Series, Vol. 6269,
  \procspie, 626931

\bibitem[{{Crane} {et~al.}(2010){Crane}, {Shectman}, {Butler}, {Thompson},
  {Birk}, {Jones}, \& {Burley}}]{crane10}
{Crane}, J.~D., {Shectman}, S.~A., {Butler}, R.~P., {et~al.} 2010, in
  \procspie, Vol. 7735, Ground-based and Airborne Instrumentation for Astronomy
  III, 773553

\bibitem[{{Crane} {et~al.}(2008){Crane}, {Shectman}, {Butler}, {Thompson}, \&
  {Burley}}]{crane08}
{Crane}, J.~D., {Shectman}, S.~A., {Butler}, R.~P., {Thompson}, I.~B., \&
  {Burley}, G.~S. 2008, in Society of Photo-Optical Instrumentation Engineers
  (SPIE) Conference Series, Vol. 7014, \procspie, 701479

\bibitem[{{Cuntz} \& {Guinan}(2016)}]{cuntz16}
{Cuntz}, M., \& {Guinan}, E.~F. 2016, \apj, 827, 79,
  \dodoi{10.3847/0004-637X/827/1/79}

\bibitem[{{Danielski} {et~al.}(2018){Danielski}, {Baudino}, {Lagage},
  {Boccaletti}, {Gastaud}, {Coulais}, \& {B{\'e}zard}}]{danielski18}
{Danielski}, C., {Baudino}, J.-L., {Lagage}, P.-O., {et~al.} 2018, \aj, 156,
  276, \dodoi{10.3847/1538-3881/aae651}

\bibitem[{Dumusque {et~al.}(2014)Dumusque, Boisse, \& Santos}]{dumusque14}
Dumusque, X., Boisse, I., \& Santos, N. 2014, The Astrophysical Journal, 796,
  132

\bibitem[{{Dumusque} {et~al.}(2017){Dumusque}, {Borsa}, {Damasso},
  {D{\'{\i}}az}, {Gregory}, {Hara}, {Hatzes}, {Rajpaul}, {Tuomi}, {Aigrain},
  {Anglada-Escud{\'e}}, {Bonomo}, {Bou{\'e}}, {Dauvergne}, {Frustagli},
  {Giacobbe}, {Haywood}, {Jones}, {Laskar}, {Pinamonti}, {Poretti}, {Rainer},
  {S{\'e}gransan}, {Sozzetti}, \& {Udry}}]{dumusque16b}
{Dumusque}, X., {Borsa}, F., {Damasso}, M., {et~al.} 2017, \aap, 598, A133,
  \dodoi{10.1051/0004-6361/201628671}

\bibitem[{Elzhov {et~al.}(2016)Elzhov, Mullen, Spiess, Bolker, Mullen, \&
  Suggests}]{elzhov16}
Elzhov, T.~V., Mullen, K.~M., Spiess, A.-N., {et~al.} 2016

\bibitem[{Feng {et~al.}(2019)Feng, Lisogorskyi, Jones, Kopeikin, Butler,
  Anglada-Escud{\'{e}}, \& Boss}]{feng19c}
Feng, F., Lisogorskyi, M., Jones, H. R.~A., {et~al.} 2019, \apjs, 244, 39,
  \dodoi{10.3847/1538-4365/ab40b6}

\bibitem[{{Feng} {et~al.}(2017{\natexlab{a}}){Feng}, {Tuomi}, \&
  {Jones}}]{feng17b}
{Feng}, F., {Tuomi}, M., \& {Jones}, H.~R.~A. 2017{\natexlab{a}}, \mnras, 470,
  4794, \dodoi{10.1093/mnras/stx1126}

\bibitem[{{Feng} {et~al.}(2017{\natexlab{b}}){Feng}, {Tuomi}, {Jones},
  {Barnes}, {Anglada-Escud{\'e}}, {Vogt}, \& {Butler}}]{feng17c}
{Feng}, F., {Tuomi}, M., {Jones}, H.~R.~A., {et~al.} 2017{\natexlab{b}}, \aj,
  154, 135, \dodoi{10.3847/1538-3881/aa83b4}

\bibitem[{{Feng} {et~al.}(2016){Feng}, {Tuomi}, {Jones}, {Butler}, \&
  {Vogt}}]{feng16}
{Feng}, F., {Tuomi}, M., {Jones}, H.~R.~A., {Butler}, R.~P., \& {Vogt}, S.
  2016, \mnras, 461, 2440, \dodoi{10.1093/mnras/stw1478}

\bibitem[{{Feng} {et~al.}(2019){Feng}, {Crane}, {Xuesong Wang}, {Teske},
  {Shectman}, {D{\'\i}az}, {Thompson}, {Jones}, \& {Butler}}]{feng19a}
{Feng}, F., {Crane}, J.~D., {Xuesong Wang}, S., {et~al.} 2019, \apjs, 242, 25,
  \dodoi{10.3847/1538-4365/ab1b16}

\bibitem[{{Feng} {et~al.}(2020){Feng}, {Butler}, {Shectman}, {Crane}, {Vogt},
  {Chambers}, {Jones}, {Xuesong Wang}, {Teske}, {Burt}, {D{\'\i}az}, \&
  {Thompson}}]{feng20a}
{Feng}, F., {Butler}, R.~P., {Shectman}, S.~A., {et~al.} 2020, \apjs, 246, 11,
  \dodoi{10.3847/1538-4365/ab5e7c}

\bibitem[{{Fischer} {et~al.}(2016){Fischer}, {Anglada-Escude}, {Arriagada},
  {Baluev}, {Bean}, {Bouchy}, {Buchhave}, {Carroll}, {Chakraborty}, {Dawson},
  {Diddams}, {Dumusque}, {Eastman}, {Endl}, {Figueira}, {Ford},
  {Foreman-Mackey}, {Fournier}, {Furesz}, {Gaudi}, {Gregory}, {Grundahl},
  {Hatsyzes}, {Hebrard}, {Herrero}, {Hogg}, {Howard}, {Johnson}, {Jorden},
  {Jurgenson}, {Latham}, {Laughlin}, {Loredo}, {Lovis}, {Mahadevan},
  {McCracken}, {Pepe}, {Perez}, {Phillips}, {Plavchan}, {Prato}, {Quirrenbach},
  {Reiners}, {Robertson}, {Santos}, {Sawyer}, {Segransan}, {Sozzetti},
  {Steinmetz}, {Szentgyorgyi}, {Udry}, {Valenti}, {Wang}, {Wittenmyer}, \&
  {Wright}}]{fischer16}
{Fischer}, D., {Anglada-Escude}, G., {Arriagada}, P., {et~al.} 2016, ArXiv
  e-prints.
\newblock \doarXiv{1602.07939}

\bibitem[{{Gaudi} {et~al.}(2020){Gaudi}, {Seager}, {Mennesson}, {Kiessling},
  {Warfield}, {Cahoy}, {Clarke}, {Domagal-Goldman}, {Feinberg}, {Guyon},
  {Kasdin}, {Mawet}, {Plavchan}, {Robinson}, {Rogers}, {Scowen}, {Somerville},
  {Stapelfeldt}, {Stark}, {Stern}, {Turnbull}, {Amini}, {Kuan}, {Martin},
  {Morgan}, {Redding}, {Stahl}, {Webb}, {Alvarez-Salazar}, {Arnold}, {Arya},
  {Balasubramanian}, {Baysinger}, {Bell}, {Below}, {Benson}, {Blais}, {Booth},
  {Bourgeois}, {Bradford}, {Brewer}, {Brooks}, {Cady}, {Caldwell}, {Calvet},
  {Carr}, {Chan}, {Cormarkovic}, {Coste}, {Cox}, {Danner}, {Davis}, {Dewell},
  {Dorsett}, {Dunn}, {East}, {Effinger}, {Eng}, {Freebury}, {Garcia}, {Gaskin},
  {Greene}, {Hennessy}, {Hilgemann}, {Hood}, {Holota}, {Howe}, {Huang}, {Hull},
  {Hunt}, {Hurd}, {Johnson}, {Kissil}, {Knight}, {Kolenz}, {Kraus}, {Krist},
  {Li}, {Lisman}, {Mandic}, {Mann}, {Marchen}, {Marrese-Reading}, {McCready},
  {McGown}, {Missun}, {Miyaguchi}, {Moore}, {Nemati}, {Nikzad}, {Nissen},
  {Novicki}, {Perrine}, {Pineda}, {Polanco}, {Putnam}, {Qureshi}, {Richards},
  {Eldorado Riggs}, {Rodgers}, {Rud}, {Saini}, {Scalisi}, {Scharf}, {Schulz},
  {Serabyn}, {Sigrist}, {Sikkia}, {Singleton}, {Shaklan}, {Smith}, {Southerd},
  {Stahl}, {Steeves}, {Sturges}, {Sullivan}, {Tang}, {Taras}, {Tesch},
  {Therrell}, {Tseng}, {Valente}, {Van Buren}, {Villalvazo}, {Warwick}, {Webb},
  {Westerhoff}, {Wofford}, {Wu}, {Woo}, {Wood}, {Ziemer}, {Arney}, {Anderson},
  {Ma{\'\i}z-Apell{\'a}niz}, {Bartlett}, {Belikov}, {Bendek}, {Cenko},
  {Douglas}, {Dulz}, {Evans}, {Faramaz}, {Feng}, {Ferguson}, {Follette},
  {Ford}, {Garc{\'\i}a}, {Geha}, {Gelino}, {G{\"o}tberg}, {Hildebrand t}, {Hu},
  {Jahnke}, {Kennedy}, {Kreidberg}, {Isella}, {Lopez}, {Marchis}, {Macri},
  {Marley}, {Matzko}, {Mazoyer}, {McCandliss}, {Meshkat}, {Mordasini},
  {Morris}, {Nielsen}, {Newman}, {Petigura}, {Postman}, {Reines}, {Roberge},
  {Roederer}, {Ruane}, {Schwieterman}, {Sirbu}, {Spalding}, {Teplitz},
  {Tumlinson}, {Turner}, {Werk}, {Wofford}, {Wyatt}, {Young}, \&
  {Zellem}}]{gaudi20}
{Gaudi}, B.~S., {Seager}, S., {Mennesson}, B., {et~al.} 2020, arXiv e-prints,
  arXiv:2001.06683.
\newblock \doarXiv{2001.06683}

\bibitem[{{Gilbert} {et~al.}(2020){Gilbert}, {Barclay}, {Schlieder},
  {Quintana}, {Hord}, {Kostov}, {Lopez}, {Rowe}, {Hoffman}, {Walkowicz},
  {Silverstein}, {Rodriguez}, {Vanderburg}, {Suissa}, {Airapetian}, {Clement},
  {Raymond}, {Mann}, {Kruse}, {Lissauer}, {Col{\'o}n}, {Kopparapu},
  {Kreidberg}, {Zieba}, {Collins}, {Quinn}, {Howell}, {Ziegler}, {Halley
  Vrijmoet}, {Adams}, {Arney}, {Boyd}, {Brande}, {Burke}, {Cacciapuoti},
  {Chance}, {Christiansen}, {Covone}, {Daylan}, {Dineen}, {Dressing}, {Essack},
  {Fauchez}, {Galgano}, {Howe}, {Kaltenegger}, {Kane}, {Lam}, {Lee}, {Lewis},
  {Logsdon}, {Mand ell}, {Monsue}, {Mullally}, {Mullally}, {Paudel},
  {Pidhorodetska}, {Plavchan}, {Ta{\~n}{\'o}n Reyes}, {Rinehart},
  {Rojas-Ayala}, {Smith}, {Stassun}, {Tenenbaum}, {Vega}, {Villanueva}, {Wolf},
  {Youngblood}, {Ricker}, {Vanderspek}, {Latham}, {Seager}, {Winn}, {Jenkins},
  {Bakos}, {Brice{\~n}o}, {Ciardi}, {Cloutier}, {Conti}, {Couperus}, {Di Sora},
  {Eisner}, {Everett}, {Gan}, {Hartman}, {Henry}, {Isopi}, {Jao}, {Jensen},
  {Law}, {Mallia}, {Matson}, {Shappee}, {Wood}, \& {Winters}}]{gilbert20}
{Gilbert}, E.~A., {Barclay}, T., {Schlieder}, J.~E., {et~al.} 2020, arXiv
  e-prints, arXiv:2001.00952.
\newblock \doarXiv{2001.00952}

\bibitem[{{Gillon} {et~al.}(2017){Gillon}, {Triaud}, {Demory}, {Jehin}, {Agol},
  {Deck}, {Lederer}, {de Wit}, {Burdanov}, {Ingalls}, {Bolmont}, {Leconte},
  {Raymond}, {Selsis}, {Turbet}, {Barkaoui}, {Burgasser}, {Burleigh}, {Carey},
  {Chaushev}, {Copperwheat}, {Delrez}, {Fernand es}, {Holdsworth}, {Kotze},
  {Van Grootel}, {Almleaky}, {Benkhaldoun}, {Magain}, \& {Queloz}}]{gillon17}
{Gillon}, M., {Triaud}, A. H.~M.~J., {Demory}, B.-O., {et~al.} 2017, \nat, 542,
  456, \dodoi{10.1038/nature21360}

\bibitem[{Haario {et~al.}(2001)Haario, Saksman, \& Tamminen}]{haario01}
Haario, H., Saksman, E., \& Tamminen, J. 2001, Bernoulli, 223

\bibitem[{Halverson {et~al.}(2016)Halverson, Terrien, Mahadevan, Roy, Bender,
  Stef{\'a}nsson, Monson, Levi, Hearty, Blake, {et~al.}}]{halverson16}
Halverson, S., Terrien, R., Mahadevan, S., {et~al.} 2016, in Ground-based and
  Airborne Instrumentation for Astronomy VI, Vol. 9908, International Society
  for Optics and Photonics, 99086P

\bibitem[{{Heller} \& {Armstrong}(2014)}]{heller14}
{Heller}, R., \& {Armstrong}, J. 2014, Astrobiology, 14, 50,
  \dodoi{10.1089/ast.2013.1088}

\bibitem[{{Jordi} {et~al.}(2010){Jordi}, {Gebran}, {Carrasco}, {de Bruijne},
  {Voss}, {Fabricius}, {Knude}, {Vallenari}, {Kohley}, \& {Mora}}]{jordi10}
{Jordi}, C., {Gebran}, M., {Carrasco}, J.~M., {et~al.} 2010, \aap, 523, A48,
  \dodoi{10.1051/0004-6361/201015441}

\bibitem[{{Jurgenson} {et~al.}(2016){Jurgenson}, {Fischer}, {McCracken},
  {Sawyer}, {Szymkowiak}, {Davis}, {Muller}, \& {Santoro}}]{jurgenson16}
{Jurgenson}, C., {Fischer}, D., {McCracken}, T., {et~al.} 2016, in Society of
  Photo-Optical Instrumentation Engineers (SPIE) Conference Series, Vol. 9908,
  Ground-based and Airborne Instrumentation for Astronomy VI, 99086T

\bibitem[{{Kane} {et~al.}(2012){Kane}, {Ciardi}, {Gelino}, \& {von
  Braun}}]{kane12}
{Kane}, S.~R., {Ciardi}, D.~R., {Gelino}, D.~M., \& {von Braun}, K. 2012,
  \mnras, 425, 757, \dodoi{10.1111/j.1365-2966.2012.21627.x}

\bibitem[{Kass \& Raftery(1995)}]{kass95}
Kass, R.~E., \& Raftery, A.~E. 1995, Journal of the american statistical
  association, 90, 773

\bibitem[{{Kay} {et~al.}(2016){Kay}, {Opher}, \& {Kornbleuth}}]{kay16}
{Kay}, C., {Opher}, M., \& {Kornbleuth}, M. 2016, \apj, 826, 195,
  \dodoi{10.3847/0004-637X/826/2/195}

\bibitem[{{Kopparapu} {et~al.}(2014){Kopparapu}, {Ramirez}, {SchottelKotte},
  {Kasting}, {Domagal-Goldman}, \& {Eymet}}]{kopparapu14}
{Kopparapu}, R.~K., {Ramirez}, R.~M., {SchottelKotte}, J., {et~al.} 2014,
  \apjl, 787, L29, \dodoi{10.1088/2041-8205/787/2/L29}

\bibitem[{{Laskar}(1997)}]{laskar97}
{Laskar}, J. 1997, \aap, 317, L75

\bibitem[{Levenberg(1944)}]{levenberg44}
Levenberg, K. 1944, Quarterly of applied mathematics, 2, 164

\bibitem[{Lomb(1976)}]{lomb76}
Lomb, N.~R. 1976, Astrophysics and space science, 39, 447

\bibitem[{{Mahadevan} {et~al.}(2012){Mahadevan}, {Ramsey}, {Bender}, {Terrien},
  {Wright}, {Halverson}, {Hearty}, {Nelson}, {Burton}, {Redman}, {Osterman},
  {Diddams}, {Kasting}, {Endl}, \& {Deshpande}}]{mahadevan12}
{Mahadevan}, S., {Ramsey}, L., {Bender}, C., {et~al.} 2012, in Society of
  Photo-Optical Instrumentation Engineers (SPIE) Conference Series, Vol. 8446,
  Ground-based and Airborne Instrumentation for Astronomy IV, 84461S

\bibitem[{Marquardt(1963)}]{marquardt63}
Marquardt, D.~W. 1963, SIAM Journal, 11, 431

\bibitem[{{Mohanty} \& {Basri}(2003)}]{mohanty03}
{Mohanty}, S., \& {Basri}, G. 2003, \apj, 583, 451, \dodoi{10.1086/345097}

\bibitem[{{Murray} \& {Dermott}(1999)}]{dermott99}
{Murray}, C.~D., \& {Dermott}, S.~F. 1999, {Solar system dynamics}

\bibitem[{Nychka {et~al.}(2018)Nychka, Furrer, Paige, Sain, \& Nychka}]{fields}
Nychka, D., Furrer, R., Paige, J., Sain, S., \& Nychka, M.~D. 2018

\bibitem[{{Oelkers} {et~al.}(2018){Oelkers}, {Rodriguez}, {Stassun}, {Pepper},
  {Somers}, {Kafka}, {Stevens}, {Beatty}, {Siverd}, {Lund}, {Kuhn}, {James}, \&
  {Gaudi}}]{oelkers18}
{Oelkers}, R.~J., {Rodriguez}, J.~E., {Stassun}, K.~G., {et~al.} 2018, \aj,
  155, 39, \dodoi{10.3847/1538-3881/aa9bf4}

\bibitem[{Pepe {et~al.}(2002)Pepe, Mayor, Rupprecht, Avila, Ballester, Beckers,
  Benz, Bertaux, Bouchy, Buzzoni, {et~al.}}]{pepe02}
Pepe, F., Mayor, M., Rupprecht, G., {et~al.} 2002, The Messenger, 110, 9

\bibitem[{{Pepe} {et~al.}(2010){Pepe}, {Cristiani}, {Rebolo Lopez}, {Santos},
  {Amorim}, {Avila}, {Benz}, {Bonifacio}, {Cabral}, {Carvas}, {Cirami},
  {Coelho}, {Comari}, {Coretti}, {De Caprio}, {Dekker}, {Delabre}, {Di
  Marcantonio}, {D'Odorico}, {Fleury}, {Garc{\'{\i}}a}, {Herreros Linares},
  {Hughes}, {Iwert}, {Lima}, {Lizon}, {Lo Curto}, {Lovis}, {Manescau},
  {Martins}, {M{\'e}gevand}, {Moitinho}, {Molaro}, {Monteiro}, {Monteiro},
  {Pasquini}, {Mordasini}, {Queloz}, {Rasilla}, {Rebord{\~a}o}, {Santana
  Tschudi}, {Santin}, {Sosnowska}, {Span{\`o}}, {Tenegi}, {Udry}, {Vanzella},
  {Viel}, {Zapatero Osorio}, \& {Zerbi}}]{pepe10}
{Pepe}, F.~A., {Cristiani}, S., {Rebolo Lopez}, R., {et~al.} 2010, in
  \procspie, Vol. 7735, Ground-based and Airborne Instrumentation for Astronomy
  III, 77350F

\bibitem[{{Perrot} {et~al.}(2018){Perrot}, {Baudoz}, {Boccaletti}, {Rousset},
  {Huby}, {Cl{\'e}net}, {Durand }, \& {Davies}}]{perrot18}
{Perrot}, C., {Baudoz}, P., {Boccaletti}, A., {et~al.} 2018, arXiv e-prints,
  arXiv:1804.01371.
\newblock \doarXiv{1804.01371}

\bibitem[{{Perruchot} {et~al.}(2008){Perruchot}, {Kohler}, {Bouchy}, {Richaud},
  {Richaud}, {Moreaux}, {Merzougui}, {Sottile}, {Hill}, {Knispel}, {Regal},
  {Meunier}, {Ilovaisky}, {Le Coroller}, {Gillet}, {Schmitt}, {Pepe}, {Fleury},
  {Sosnowska}, {Vors}, {M{\'e}gevand}, {Blanc}, {Carol}, {Point}, {Laloge}, \&
  {Brunel}}]{perruchot08}
{Perruchot}, S., {Kohler}, D., {Bouchy}, F., {et~al.} 2008, Society of
  Photo-Optical Instrumentation Engineers (SPIE) Conference Series, Vol. 7014,
  {The SOPHIE spectrograph: design and technical key-points for high throughput
  and high stability}, 70140J

\bibitem[{{Perryman} {et~al.}(2014){Perryman}, {Hartman}, {Bakos}, \&
  {Lindegren}}]{perryman14}
{Perryman}, M., {Hartman}, J., {Bakos}, G.~{\'A}., \& {Lindegren}, L. 2014,
  \apj, 797, 14, \dodoi{10.1088/0004-637X/797/1/14}

\bibitem[{{Quirrenbach} {et~al.}(2010){Quirrenbach}, {Amado}, {Mandel},
  {Caballero}, {Ribas}, {Reiners}, {Mundt}, \& {CARMENES
  Consortium}}]{quirrenbach09}
{Quirrenbach}, A., {Amado}, P.~J., {Mandel}, H., {et~al.} 2010, in Astronomical
  Society of the Pacific Conference Series, Vol. 430, Pathways Towards
  Habitable Planets, ed. V.~{Coud{\'e} du Foresto}, D.~M. {Gelino}, \&
  I.~{Ribas}, 521

\bibitem[{{Ribas} {et~al.}(2018){Ribas}, {Tuomi}, {Reiners}, {Butler},
  {Morales}, {Perger}, {Dreizler}, {Rodr{\'\i}guez-L{\'o}pez}, {Gonz{\'a}lez
  Hern{\'a}ndez}, {Rosich}, {Feng}, {Trifonov}, {Vogt}, {Caballero}, {Hatzes},
  {Herrero}, {Jeffers}, {Lafarga}, {Murgas}, {Nelson}, {Rodr{\'\i}guez},
  {Strachan}, {Tal-Or}, {Teske}, {Toledo-Padr{\'o}n}, {Zechmeister},
  {Quirrenbach}, {Amado}, {Azzaro}, {B{\'e}jar}, {Barnes}, {Berdi{\~n}as},
  {Burt}, {Coleman}, {Cort{\'e}s-Contreras}, {Crane}, {Engle}, {Guinan},
  {Haswell}, {Henning}, {Holden}, {Jenkins}, {Jones}, {Kaminski}, {Kiraga},
  {K{\"u}rster}, {Lee}, {L{\'o}pez-Gonz{\'a}lez}, {Montes}, {Morin}, {Ofir},
  {Pall{\'e}}, {Rebolo}, {Reffert}, {Schweitzer}, {Seifert}, {Shectman},
  {Staab}, {Street}, {Su{\'a}rez Mascare{\~n}o}, {Tsapras}, {Wang}, \&
  {Anglada-Escud{\'e}}}]{ribas18}
{Ribas}, I., {Tuomi}, M., {Reiners}, A., {et~al.} 2018, \nat, 563, 365,
  \dodoi{10.1038/s41586-018-0677-y}

\bibitem[{Ripley {et~al.}(2013)Ripley, Venables, Bates, Hornik, Gebhardt,
  Firth, \& Ripley}]{ripley13}
Ripley, B., Venables, B., Bates, D.~M., {et~al.} 2013, Cran R, 538

\bibitem[{{Robotham}(2016)}]{robotham16}
{Robotham}, A.~S.~G. 2016, {magicaxis: Pretty scientific plotting with
  minor-tick and log minor-tick support}, Astrophysics Source Code Library.
\newblock \doeprint{1604.004}

\bibitem[{Scargle(1982)}]{scargle82}
Scargle, J.~D. 1982, The Astrophysical Journal, 263, 835

\bibitem[{{Schwab} {et~al.}(2016){Schwab}, {Rakich}, {Gong}, {Mahadevan},
  {Halverson}, {Roy}, {Terrien}, {Robertson}, {Hearty}, {Levi}, {Monson},
  {Wright}, {McElwain}, {Bender}, {Blake}, {St{\"u}rmer}, {Gurevich},
  {Chakraborty}, \& {Ramsey}}]{schwab16}
{Schwab}, C., {Rakich}, A., {Gong}, Q., {et~al.} 2016, in \procspie, Vol. 9908,
  Ground-based and Airborne Instrumentation for Astronomy VI, 99087H

\bibitem[{Shields {et~al.}(2016)Shields, Ballard, \& Johnson}]{shields16}
Shields, A.~L., Ballard, S., \& Johnson, J.~A. 2016, Physics Reports, 663, 1

\bibitem[{{Soubiran} {et~al.}(2018){Soubiran}, {Jasniewicz}, {Chemin},
  {Zurbach}, {Brouillet}, {Panuzzo}, {Sartoretti}, {Katz}, {Le Campion},
  {Marchal}, {Hestroffer}, {Th{\'e}venin}, {Crifo}, {Udry}, {Cropper},
  {Seabroke}, {Viala}, {Benson}, {Blomme}, {Jean-Antoine}, {Huckle}, {Smith},
  {Baker}, {Damerdji}, {Dolding}, {Fr{\'e}mat}, {Gosset}, {Guerrier}, {Guy},
  {Haigron}, {Jan{\ss}en}, {Plum}, {Fabre}, {Lasne}, {Pailler}, {Panem},
  {Riclet}, {Royer}, {Tauran}, {Zwitter}, {Gueguen}, \& {Turon}}]{soubiran18}
{Soubiran}, C., {Jasniewicz}, G., {Chemin}, L., {et~al.} 2018, \aap, 616, A7,
  \dodoi{10.1051/0004-6361/201832795}

\bibitem[{Spiegelhalter {et~al.}(2002)Spiegelhalter, Best, Carlin, \& Van
  Der~Linde}]{spiegelhalter02}
Spiegelhalter, D.~J., Best, N.~G., Carlin, B.~P., \& Van Der~Linde, A. 2002,
  Journal of the Royal Statistical Society: Series B (Statistical Methodology),
  64, 583

\bibitem[{{Stassun} {et~al.}(2019){Stassun}, {Oelkers}, {Paegert}, {Torres},
  {Pepper}, {De Lee}, {Collins}, {Latham}, {Muirhead}, {Chittidi},
  {Rojas-Ayala}, {Fleming}, {Rose}, {Tenenbaum}, {Ting}, {Kane}, {Barclay},
  {Bean}, {Brassuer}, {Charbonneau}, {Ge}, {Lissauer}, {Mann}, {McLean},
  {Mullally}, {Narita}, {Plavchan}, {Ricker}, {Sasselov}, {Seager}, {Sharma},
  {Shiao}, {Sozzetti}, {Stello}, {Vanderspek}, {Wallace}, \&
  {Winn}}]{stassun19}
{Stassun}, K.~G., {Oelkers}, R.~J., {Paegert}, M., {et~al.} 2019, \aj, 158,
  138, \dodoi{10.3847/1538-3881/ab3467}

\bibitem[{{Suzuki} {et~al.}(2016){Suzuki}, {Bennett}, {Sumi}, {Bond}, {Rogers},
  {Abe}, {Asakura}, {Bhattacharya}, {Donachie}, {Freeman}, {Fukui}, {Hirao},
  {Itow}, {Koshimoto}, {Li}, {Ling}, {Masuda}, {Matsubara}, {Muraki},
  {Nagakane}, {Onishi}, {Oyokawa}, {Rattenbury}, {Saito}, {Sharan}, {Shibai},
  {Sullivan}, {Tristram}, {Yonehara}, \& {MOA Collaboration}}]{suzuki16}
{Suzuki}, D., {Bennett}, D.~P., {Sumi}, T., {et~al.} 2016, \apj, 833, 145,
  \dodoi{10.3847/1538-4357/833/2/145}

\bibitem[{{Tal-Or} {et~al.}(2019){Tal-Or}, {Trifonov}, {Zucker}, {Mazeh}, \&
  {Zechmeister}}]{talor19}
{Tal-Or}, L., {Trifonov}, T., {Zucker}, S., {Mazeh}, T., \& {Zechmeister}, M.
  2019, \mnras, 484, L8, \dodoi{10.1093/mnrasl/sly227}

\bibitem[{{Tamura} {et~al.}(2012){Tamura}, {Suto}, {Nishikawa}, {Kotani},
  {Sato}, {Aoki}, {Usuda}, {Kurokawa}, {Kashiwagi}, {Nishiyama}, {Ikeda},
  {Hall}, {Hodapp}, {Hashimoto}, {Morino}, {Inoue}, {Mizuno}, {Washizaki},
  {Tanaka}, {Suzuki}, {Kwon}, {Suenaga}, {Oh}, {Narita}, {Kokubo}, {Hayano},
  {Izumiura}, {Kambe}, {Kudo}, {Kusakabe}, {Ikoma}, {Hori}, {Omiya}, {Genda},
  {Fukui}, {Fujii}, {Guyon}, {Harakawa}, {Hayashi}, {Hidai}, {Hirano},
  {Kuzuhara}, {Machida}, {Matsuo}, {Nagata}, {Ohnuki}, {Ogihara}, {Oshino},
  {Suzuki}, {Takami}, {Takato}, {Takahashi}, {Tachinami}, \&
  {Terada}}]{tamura12}
{Tamura}, M., {Suto}, H., {Nishikawa}, J., {et~al.} 2012, in Society of
  Photo-Optical Instrumentation Engineers (SPIE) Conference Series, Vol. 8446,
  Ground-based and Airborne Instrumentation for Astronomy IV, 84461T

\bibitem[{Tang {et~al.}(2019)Tang, Rodgers, Creager, Krist, McGuire, Patterson,
  Rud, Shi, \& Zhao}]{tang19}
Tang, H., Rodgers, M., Creager, B., {et~al.} 2019, in Techniques and
  Instrumentation for Detection of Exoplanets IX, Vol. 11117, International
  Society for Optics and Photonics, 111170C

\bibitem[{Tarter {et~al.}(2007)Tarter, Backus, Mancinelli, Aurnou, Backman,
  Basri, Boss, Clarke, Deming, Doyle, {et~al.}}]{tarter07}
Tarter, J.~C., Backus, P.~R., Mancinelli, R.~L., {et~al.} 2007, Astrobiology,
  7, 30

\bibitem[{{Trifonov} {et~al.}(2020){Trifonov}, {Tal-Or}, {Zechmeister},
  {Kaminski}, {Zucker}, \& {Mazeh}}]{trifonov20}
{Trifonov}, T., {Tal-Or}, L., {Zechmeister}, M., {et~al.} 2020, \aap, 636, A74,
  \dodoi{10.1051/0004-6361/201936686ARXIV: 2001.05942OPEN}

\bibitem[{Tuomi \& Anglada-Escud{\'e}(2013)}]{tuomi13}
Tuomi, M., \& Anglada-Escud{\'e}, G. 2013, Astronomy \& Astrophysics, 556, A111

\bibitem[{{Van Eylen} {et~al.}(2019){Van Eylen}, {Albrecht}, {Huang},
  {MacDonald}, {Dawson}, {Cai}, {Foreman-Mackey}, {Lundkvist}, {Silva Aguirre},
  {Snellen}, \& {Winn}}]{VanEylen18}
{Van Eylen}, V., {Albrecht}, S., {Huang}, X., {et~al.} 2019, \aj, 157, 61,
  \dodoi{10.3847/1538-3881/aaf22f}

\bibitem[{Vogt {et~al.}(1994)Vogt, Allen, Bigelow, Bresee, Brown, Cantrall,
  Conrad, Couture, Delaney, Epps, Hilyard, Hilyard, Horn, Jern, Kanto, Keane,
  Kibrick, Lewis, Osborne, Pardeilhan, Pfister, Ricketts, Robinson, Stover,
  Tucker, Ward, \& Wei}]{vogt94}
Vogt, S.~S., Allen, S.~L., Bigelow, B.~C., {et~al.} 1994, in Instrumentation in
  Astronomy VIII, ed. D.~L. Crawford \& E.~R. Craine, Vol. 2198, International
  Society for Optics and Photonics (SPIE), 362 -- 375.
\newblock \url{https://doi.org/10.1117/12.176725}

\bibitem[{{Vogt} {et~al.}(2014){Vogt}, {Radovan}, {Kibrick}, {Butler},
  {Alcott}, {Allen}, {Arriagada}, {Bolte}, {Burt}, {Cabak}, {Chloros},
  {Cowley}, {Deich}, {Dupraw}, {Earthman}, {Epps}, {Faber}, {Fischer}, {Gates},
  {Hilyard}, {Holden}, {Johnston}, {Keiser}, {Kanto}, {Katsuki}, {Laiterman},
  {Lanclos}, {Laughlin}, {Lewis}, {Lockwood}, {Lynam}, {Marcy}, {McLean},
  {Miller}, {Misch}, {Peck}, {Pfister}, {Phillips}, {Rivera}, {Sandford},
  {Saylor}, {Stover}, {Thompson}, {Walp}, {Ward}, {Wareham}, {Wei}, \&
  {Wright}}]{vogt14}
{Vogt}, S.~S., {Radovan}, M., {Kibrick}, R., {et~al.} 2014, Publications of the
  Astronomical Society of the Pacific, 126, 359, \dodoi{10.1086/676120}

\bibitem[{{Wenger} {et~al.}(2000){Wenger}, {Ochsenbein}, {Egret}, {Dubois},
  {Bonnarel}, {Borde}, {Genova}, {Jasniewicz}, {Lalo{\"e}}, {Lesteven}, \&
  {Monier}}]{wenger00}
{Wenger}, M., {Ochsenbein}, F., {Egret}, D., {et~al.} 2000, \aaps, 143, 9,
  \dodoi{10.1051/aas:2000332}

\bibitem[{{West} {et~al.}(2015){West}, {Weisenburger}, {Irwin},
  {Berta-Thompson}, {Charbonneau}, {Dittmann}, \& {Pineda}}]{west15}
{West}, A.~A., {Weisenburger}, K.~L., {Irwin}, J., {et~al.} 2015, \apj, 812, 3,
  \dodoi{10.1088/0004-637X/812/1/3}

\bibitem[{Wright \& Eastman(2014)}]{wright14}
Wright, J., \& Eastman, J. 2014, Publications of the Astronomical Society of
  the Pacific, 126, 838

\bibitem[{{Zechmeister} {et~al.}(2018){Zechmeister}, {Reiners}, {Amado},
  {Azzaro}, {Bauer}, {B{\'e}jar}, {Caballero}, {Guenther}, {Hagen}, {Jeffers},
  {Kaminski}, {K{\"u}rster}, {Launhardt}, {Montes}, {Morales}, {Quirrenbach},
  {Reffert}, {Ribas}, {Seifert}, {Tal-Or}, \& {Wolthoff}}]{zechmeister18}
{Zechmeister}, M., {Reiners}, A., {Amado}, P.~J., {et~al.} 2018, \aap, 609,
  A12, \dodoi{10.1051/0004-6361/201731483}

\bibitem[{{Zechmeister} {et~al.}(2019){Zechmeister}, {Dreizler}, {Ribas},
  {Reiners}, {Caballero}, {Bauer}, {B{\'e}jar}, {Gonz{\'a}lez-Cuesta},
  {Herrero}, {Lalitha}, {L{\'o}pez-Gonz{\'a}lez}, {Luque}, {Morales},
  {Pall{\'e}}, {Rodr{\'\i}guez}, {Rodr{\'\i}guez L{\'o}pez}, {Tal-Or},
  {Anglada-Escud{\'e}}, {Quirrenbach}, {Amado}, {Abril}, {Aceituno},
  {Aceituno}, {Alonso-Floriano}, {Ammler-von Eiff}, {Antona Jim{\'e}nez},
  {Anwand-Heerwart}, {Arroyo-Torres}, {Azzaro}, {Baroch}, {Barrado},
  {Becerril}, {Ben{\'\i}tez}, {Berdi{\~n}as}, {Bergond}, {Bluhm},
  {Brinkm{\"o}ller}, {del Burgo}, {Calvo Ortega}, {Cano}, {Cardona
  Guill{\'e}n}, {Carro}, {C{\'a}rdenas V{\'a}zquez}, {Casal},
  {Casasayas-Barris}, {Casanova}, {Chaturvedi}, {Cifuentes}, {Claret},
  {Colom{\'e}}, {Cort{\'e}s-Contreras}, {Czesla}, {D{\'\i}ez-Alonso}, {Dorda},
  {Fern{\'a}ndez}, {Fern{\'a}ndez-Mart{\'\i}n}, {Fuhrmeister}, {Fukui},
  {Galad{\'\i}-Enr{\'\i}quez}, {Gallardo Cava}, {Garcia de la Fuente},
  {Garcia-Piquer}, {Garc{\'\i}a Vargas}, {Gesa}, {G{\'o}ngora Rueda},
  {Gonz{\'a}lez-{\'A}lvarez}, {Gonz{\'a}lez Hern{\'a}ndez},
  {Gonz{\'a}lez-Peinado}, {Gr{\"o}zinger}, {Gu{\`a}rdia}, {Guijarro}, {de
  Guindos}, {Hatzes}, {Hauschildt}, {Hedrosa}, {Helmling}, {Henning},
  {Hermelo}, {Hern{\'a}ndez Arabi}, {Hern{\'a}ndez Casta{\~n}o}, {Hern{\'a}ndez
  Otero}, {Hintz}, {Huke}, {Huber}, {Jeffers}, {Johnson}, {de Juan},
  {Kaminski}, {Kemmer}, {Kim}, {Klahr}, {Klein}, {Kl{\"u}ter}, {Klutsch},
  {Kossakowski}, {K{\"u}rster}, {Labarga}, {Lafarga}, {Llamas}, {Lamp{\'o}n},
  {Lara}, {Launhardt}, {L{\'a}zaro}, {Lodieu}, {L{\'o}pez del Fresno},
  {L{\'o}pez-Puertas}, {L{\'o}pez Salas}, {L{\'o}pez-Santiago}, {Mag{\'a}n
  Madinabeitia}, {Mall}, {Mancini}, {Mand el}, {Marfil}, {Mar{\'\i}n Molina},
  {Maroto Fern{\'a}ndez}, {Mart{\'\i}n}, {Mart{\'\i}n-Fern{\'a}ndez},
  {Mart{\'\i}n-Ruiz}, {Marvin}, {Mirabet}, {Monta{\~n}{\'e}s-Rodr{\'\i}guez},
  {Montes}, {Moreno-Raya}, {Nagel}, {Naranjo}, {Narita}, {Nortmann}, {Nowak},
  {Ofir}, {Oshagh}, {Panduro}, {Parviainen}, {Pascual}, {Passegger}, {Pavlov},
  {Pedraz}, {P{\'e}rez-Calpena}, {P{\'e}rez Medialdea}, {Perger}, {Perryman},
  {Rabaza}, {Ram{\'o}n Ballesta}, {Rebolo}, {Redondo}, {Reffert}, {Reinhardt},
  {Rhode}, {Rix}, {Rodler}, {Rodr{\'\i}guez Trinidad}, {Rosich}, {Sadegi},
  {S{\'a}nchez-Blanco}, {S{\'a}nchez Carrasco}, {S{\'a}nchez-L{\'o}pez},
  {Sanz-Forcada}, {Sarkis}, {Sarmiento}, {Sch{\"a}fer}, {Schmitt},
  {Sch{\"o}fer}, {Schweitzer}, {Seifert}, {Shulyak}, {Solano}, {Sota}, {Stahl},
  {Stock}, {Strachan}, {Stuber}, {St{\"u}rmer}, {Su{\'a}rez}, {Tabernero},
  {Tala Pinto}, {Trifonov}, {Veredas}, {Vico Linares}, {Vilardell}, {Wagner},
  {Wolthoff}, {Xu}, {Yan}, \& {Zapatero Osorio}}]{zechmeister19}
{Zechmeister}, M., {Dreizler}, S., {Ribas}, I., {et~al.} 2019, \aap, 627, A49,
  \dodoi{10.1051/0004-6361/201935460}

\bibitem[{{Zhu} \& {Wu}(2018)}]{zhu18}
{Zhu}, W., \& {Wu}, Y. 2018, \aj, 156, 92, \dodoi{10.3847/1538-3881/aad22a}

\end{thebibliography}
\end{document}